\pdfoutput=1   

\documentclass[pra,onecolumn,showpacs,preprintnumbers,amsmath,amssymb,longbibliography]{revtex4-1} 

\usepackage{amssymb}
\usepackage{amsmath}
\usepackage{graphicx}
\usepackage[fleqn]{mathtools}
\usepackage{siunitx}
\usepackage{xcolor}
\usepackage{hyperref}

\renewcommand{\vec}[1]{\mathbf{{#1}}}                         
\newcommand{\pvec}[1]{\mathbf{{#1}}_\parallel }
\newcommand{\vecUnit}[1]{\mathbf{\hat{#1}}}                         
\newcommand{\pvecUnit}[1]{\mathbf{\hat{#1}}_\parallel }
\newcommand{\nn}{\nonumber}
 \newcommand{\imu}{\mathrm{i}}
\newcommand{\dint}[2][]{\!\mathrm{d}^{#1}#2\,}                 
\newcommand{\dfint}[3][]{\! \frac{\mathrm{d}^{#1}#2}{#3}\,}

\renewcommand{\Re}{\mathrm{Re}\,}
\renewcommand{\Im}{\mathrm{Im}\,}

\DeclareMathOperator{\sgn}{sgn}

\newcommand{\bqe}{\begin{eqnarray}}
\newcommand{\eqe}{\end{eqnarray}}
\newcommand{\bseq}{\begin{subequations}}
\newcommand{\eseq}{\end{subequations}}

\newcommand{\zxp}{\zeta({\bf x}_{\|})}
\newcommand{\zxpp}{\zeta({\bf x}\, '\!\!_{\|} )}
\newcommand{\bkp}{{\bf k}_{\|}}
\newcommand{\bqp}{{\bf q}_{\|}}
\newcommand{\bup}{{\bf u}_{\|}}
\newcommand{\bpp}{{\bf p}_{\|}}

\newcommand{\bal}{\begin{align}}
\newcommand{\eal}{\end{align}}
\newcommand{\qp}{q_{\|}}
\newcommand{\pp}{p_{\|}}

\newcommand{\bQp}{{\bf Q}_{\|}}
\newcommand{\bQpp}{{\bf Q}\, '\!\!_{\|}}
\newcommand{\bxp}{{\bf x}_{\|}}

\newcommand{\bzero}{{\bf 0}}
\newcommand{\bxpp}{{\bf x}\, '\!\!_{\|}}

\newcommand{\p}{\partial }

\newcommand{\e}{\varepsilon}

\newcommand{\kp}{k_{\|}}
\newcommand{\up}{u_{\|}}
\newcommand{\hbkp}{\hat{\bf k}_{\|}}

\newcommand{\zhq}{\hat{\zeta}({\bf Q}_{\|})}
\newcommand{\zhqp}{\hat{\zeta}({\bf Q}\, '\!\!_{\|})}

\newcommand{\hbpp}{\hat{\bf p}_{\|}}
\newcommand{\w}{\omega }

\newcommand{\hbqp}{\hat{{\bf q}}_{\|}}

\newcommand{\la}{\langle}
\newcommand{\ra}{\rangle}

\newcommand{\xp}{x_{\|}}

\graphicspath{{Figures/},{Figures.New/}}         

\begin{document} 


\title{Determination of the normalized surface height autocorrelation function of a two-dimensional randomly rough dielectric surface by the inversion of light scattering data}

\author{I. Simonsen$^{1}$}
\email{ingve.simonsen@ntnu.no}
\author{\O.S. Hetland$^1$} 
\author{J.B. Kryvi$^1$} 
\author{A.A. Maradudin$^2$}
\email{aamaradu@uci.edu}

\affiliation{$^1$Department of Physics, Norwegian University of Science and Technology, NO-7491 Trondheim, Norway}
\affiliation{$^2$Department of Physics and Astronomy, University of California, Irvine CA 92697, U.S.A.}

\date{\today}

\begin{abstract}
  An expression is obtained on the basis of phase perturbation theory for the contribution to the mean differential reflection coefficient from the in-plane co-polarized component of the light scattered diffusely from a two-dimensional randomly rough dielectric surface when the latter is illuminated by s-polarized light. This result forms the basis for an approach to inverting experimental light scattering data to obtain the normalized surface height autocorrelation function of the surface. Several parametrized forms of this correlation function, and the minimization of a cost function with respect to the parameters defining these representations, are used in the inversion scheme. This approach also yields the rms height of the surface roughness, and the dielectric constant of the dielectric substrate if it is not known in advance. The input data used in validating this inversion consists of computer simulation results for surfaces defined by exponential and Gaussian surface height correlation functions, without and with the addition of multiplicative noise, for a single or multiple  angles of incidence. The reconstructions obtained by this approach are quite accurate for weakly rough surfaces, and the proposed inversion scheme is computationally efficient.   
\end{abstract}


\maketitle

\section{Introduction}

Statistical information about the roughness of a surface is contained in its rms height and in its normalized surface height autocorrelation function. Efforts to obtain these properties of rough surfaces from measurements of light scattered into the far field from them are of interest because 
of the contactless nature of this approach, and because measurements in the far field are easier to carry out than measurements in the near field. 

This problem has been studied in the past by several authors. In the case of a one-dimensional randomly rough dielectric surface, Chakrabarti {\it et al.}~\cite{1} inverted by a Fourier transformation an expression for the contribution to the mean differential reflection coefficient, obtained in the Kirchhoff  approximation, from light scattered diffusely from the surface. The incident light was s-polarized and the plane of incidence was perpendicular to the generators of the surface. Good agreement with numerically generated scattering data was obtained for weakly rough surfaces. 

The case of a two-dimensional randomly rough surface was studied by Chandley~\cite{2}, and by Marx and Vorburger~\cite{3}. Chandley used scalar diffraction theory and a thin random phase screen approximation~\cite{4} to model the interaction of light with the randomly rough surface. A thin 
phase screen may be regarded as a layer of negligible thickness that alters the phase of the wave scattered from it but does not change its magnitude. It is derived from simple optical path length and geometrical optics arguments. He used the angular dependence of the mean intensity of the scattered light in the far field as the experimental quantity to be inverted. The nature of his scattering model allowed this inversion to be carried out by means of a Fourier transformation. The dielectric constant of the scattering medium does not appear explicitly in Chandley's theory which means that it is impossible to use it to recover the dielectric constant of the scattering medium from the experimental scattering data if it is not known in advance.

In their study of this problem, Marx and Vorburger applied the Kirchhoff approximation for the scattering of a scalar plane wave from a two-dimensional randomly rough perfectly conducting surface to obtain the mean intensity of the scattered field. The determination of the rms height of the surface and the normalized surface height autocorrelation function was achieved by assuming an expression for the latter function of a particular analytic form and by the determination of the parameters defining it by a least squares fit of the theoretical mean intensity to the experimental result. 

In contrast to these studies, in this paper we present an approach to the determination of the rms height and the normalized surface height autocorrelation function of a two-dimensional randomly  rough penetrable surface, in particular a dielectric surface, from the inversion of optical scattering data. It is based on a vector theory of rough surface scattering rather than on a scalar theory, namely phase perturbation theory~\cite{5}. The dielectric constant of the medium is taken into account in this approach. This version of rough surface scattering theory was chosen in this study because in a recent comparison between experimental data and the predictions of three perturbation theories for the scattering of electromagnetic radiation from two-dimensional randomly rough metal surfaces, it produced the best results~\cite{5}. We expect it to be equally accurate in describing the scattering of visible light from a two-dimensional randomly  rough dielectric surface. Specifically, we use the expression for the contribution to the mean differential reflection coefficient from the in-plane, co-polarized component of the light scattered incoherently when the dielectric surface is illuminated at normal or non-normal incidence by s-polarized light.

This expression is evaluated with the use of an expression for the normalized surface height autocorrelation function  that contains adjustable parameters. The values of these parameters are then determined by a least squares fit of the resulting expression to the corresponding experimental scattering data. The reconstruction of the parameters is performed using scattering data for both a single and several angles of incidence, and the sensitivity to noise of the reconstructed parameters is investigated. We note that the contribution to the mean differential reflection coefficient from the in-plane co-polarized component of the light scattered incoherently when the surface is illuminated by p-polarized light can also be used for this purpose. However, the expression one works with to effect this inversion is somewhat simpler in s polarization than in p polarization. In addition, there is no Brewster effect in s polarization, so that a smoother function of the scattering angle is being inverted in s polarization than in p polarization. It is for these reasons that we have chosen to work with s polarization.

\smallskip
This paper is organized as follows: First, the scattering system is presented [Sec.~\ref{Sec:geomtry}] followed by elements of scattering theory [Sec.~\ref{Sec:theory}] that will be useful for the subsequent discussion. Then in Sec.~\ref{Sec:Inversion}, we present the inversion scheme that will be used to reconstruct the surface height autocorrelation function. The results obtained by the use of this procedure are presented in Sec.~\ref{sec:Results} for a set of different correlation functions and scattering geometries. Section~\ref{Sec:Conclusions} presents discussions of these results  and the conclusions that can be drawn from this study. The paper ends with an Appendix detailing the derivation of expressions, central to the present work,  for the first few moments of the scattering matrix for s-to-s scattering obtained on the basis of phase perturbation theory.

\section{The Physical System Studied} 
\label{Sec:geomtry}

%
\begin{figure}
  \includegraphics{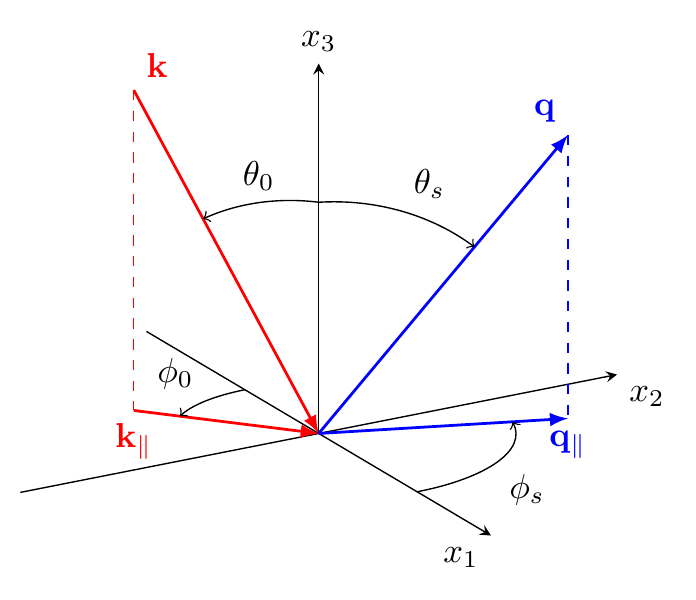}
  \caption{Schematics of the scattering geometry considered in this work.}
  \label{fig:geometry}
\end{figure}

The physical system we study in this paper consists of vacuum in the region $x_3 > \zxp$, and a dielectric medium, characterized by a dielectric constant $\varepsilon$ that is real, positive and frequency independent, in the region $x_3 < \zxp$~[Fig.~\ref{fig:geometry}]. Here $\pvec{x} = (x_1, x_2, 0)$ is a position vector in the plane $x_3  =0$. The surface profile function $\zxp$ is assumed to be a single-valued function of $\bxp$ that is differentiable with respect to $x_1$ and $x_2$. It is also assumed to constitute a stationary, zero-mean, isotropic, Gaussian random process defined by 

\begin{subequations}
  \label{eq:1}
  \begin{align}
    \label{eq:1a}
    \langle \zxp \zxpp \rangle &= \delta^2 W(|\bxp - \bxpp|) \\
    \label{eq:1b}
    \langle \zeta^2({\bf x}_{\|}) \rangle &= \delta^2,
  \end{align}
\end{subequations}
where the angle brackets denote an average over the ensemble of realizations of $\zxp$, $\delta$ is the rms height of the surface, and $W(|\bxp|)$ is  the {\it normalized surface height autocorrelation function}, with the property that $W({\bf 0}) = 1$. 

The surface profile function has a Fourier integral representation,
\begin{align}
  \label{eq:2}
  \zxp &= \int \dfint[2]{Q_{\parallel}}{(2 \pi)^2} \zhq \exp(\imu \bQp \cdot \bxp ),
\end{align}
where $\bQp = (Q_1, Q_2, 0)$ is a two-dimensional wave vector so that
\bseq
\label{eq:3a}
\begin{align}
  \zhq &= \int \dint[2]{x_{\parallel}} \zxp \mathrm{exp}(-\imu \bQp \cdot \bxp ).
\end{align}
We also introduce the notation
\label{eq:3b}
\begin{align}
  \hat{\zeta}^{(n)}(\pvec{Q}) &= \int \dint[2]{x_{\parallel}} \zeta^n({\bf x}_{\|}) \exp(-\imu \bQp \cdot \bxp ).
\end{align}
\eseq

The Fourier coefficient $\zhq$ is also a zero-mean Gaussian random process defined by 
\begin{align}
  \label{eq:4}
  \left< \zhq \zhqp \right> &=(2 \pi)^2 \delta(\bQp + \bQpp) \, \delta^2 g(|\bQp|),
\end{align}
where $g(|\bQp|)$, the {\it power spectrum} of the surface roughness, is defined by
\begin{align}
  \label{eq:5}
  g(|\bQp|) &= \int \dint[2]{x_{\parallel}} W(|\bxp |) \, \exp(- \imu \bQp \cdot \bxp ).
\end{align}

It follows from Eqs.~\eqref{eq:1} and \eqref{eq:5} that $g(|\bQp|)$ is normalized to unity,
\begin{align}
\label{eq:6}
\int \frac{d^2 Q_{\parallel}}{(2 \pi)^2} g(|\bQp|) &= 1.
\end{align}

\section{Scattering Theory}
\label{Sec:theory}

The surface $x_3 = \zxp$ is illuminated from the vacuum by an electromagnetic field of frequency $\omega$. 
The electric field in the vacuum above the surface is the sum of an incident and a scattered field, ${\bf E}({\bf x}; t) = {\bf E}^{(i)}({\bf x};t) 
+ {\bf E}^{(s)}({\bf x};t)$, where 
\begin{subequations}
  \label{eq:7}
  \begin{align}
    \label{eq:7a}
    \vec{E}^{(i)}(\vec{x};t) & = \Big\{
                                   -\frac{c}{\omega} \left[ \pvecUnit{k} \alpha_0(\kp) + \vecUnit{x}_3 \kp \right] B_p(\bkp)   
                                   + ( \vecUnit{x}_3 \times \pvecUnit{k} )B_s(\bkp) 
                            \Big\}\,
                            \exp\left[ \imu (\bkp - \vecUnit{x}_3\alpha_0(\kp)) \cdot \vec{x} - \imu \omega t\right]
  \end{align}
  \begin{align}
    \label{eq:7b}
    \vec{E}^{(s)}(\vec{x};t) & = \int \dfint[2]{\qp}{(2 \pi)^2}
        \Big\{ 
              \frac{c}{\omega} \left[ \pvecUnit{q} \alpha_0(\qp) - \vecUnit{x}_3 \qp \right] A_p(\bqp)   
              + (\vecUnit{x}_3 \times \pvecUnit{q} )A_s(\bqp) 
        \Big\} \,
        \exp \left[ \imu(\bqp + \vecUnit{x}_3\alpha_0(\qp)) \cdot \vec{x} - \imu\omega t\right].
   \end{align}
\end{subequations}
The subscripts p and s denote the p-polarized (TM) and s-polarized (TE) components of each of these fields, respectively. The function $\alpha_0(\qp)$ in
Eqs.~\eqref{eq:7} is 
defined as
\begin{align}
  \label{eq:def-alpha0}
  \alpha_0(\qp) &= \left[\left(\frac{\w}{c}\right)^2 - \qp^2 \right]^{1/2} 
     \qquad \quad \Re\alpha_0(\qp) > 0, \,\, \Im\alpha_0(\qp) > 0. 
\end{align}
Maxwell's equations imply linear relations between $A_\alpha(\bqp)$ and $B_\beta(\bqp)$, which we write in the form $(\alpha = p, s, \beta = p, s)$
\begin{align} 
  \label{eq:8}
  A_\alpha(\bqp) & = \sum\limits_\beta R_{\alpha \beta}(\bqp|\bkp) B_\beta(\bkp).
\end{align}
The scattering amplitudes $\{R_{\alpha \beta}(\bqp|\bkp)\}$ play a significant role in the present theory because the mean differential reflection coefficient  is given in terms of them. The differential reflection coefficient $\left(\partial R_{\alpha \beta}(\bqp|\bkp)/ \partial \Omega_s \right)$ is defined such that $\left(\partial R_{\alpha \beta}(\bqp|\bkp)/ \partial \Omega_s \right)d\Omega_s$ is the fraction of the total time -averaged flux in an incident field of $\beta$ polarization the projection of whose wave vector on the mean scattering plane is $\bkp$, that is scattered into a field of $\alpha$ polarization, the projection  of whose wave vector on the mean scattering plane is $\bqp$, within an element of solid angle $d\Omega_s$ about the scattering direction defined by the polar  and azimuthal angles $(\theta_s, \phi_s)$. It is given by~\cite{our1,review}
\begin{align}
  \label{eq:9}
  \frac{\partial R_{\alpha \beta}(\bqp|\bkp)}{\partial \Omega_s} 
  &= 
  \frac{1}{S}\left(\frac{\omega}{2 \pi c}\right)^2 \frac{\cos^2 \theta_s}{\cos \theta_0}\left|R_{\alpha \beta}(\bqp|\bkp)\right|^2, 
\end{align}
with [see Fig.~\ref{fig:geometry}]
\begin{subequations}
  \label{eq:9-1}
  \begin{align}
    \pvec{k} &= \frac{\omega}{c} \sin \theta_0(\cos \phi_0, \sin \phi_0, 0)
    \label{eq:9-1a}
    \\
    \pvec{q} &= \frac{\omega}{c} \sin \theta_s(\cos \phi_s, \sin \phi_s, 0),
    \label{eq:9-1b} 
  \end{align}
\end{subequations}
where $(\theta_0, \phi_0)$  and $(\theta_s, \phi_s)$ are the polar and azimuthal angles of incidence and scattering, respectively. $S$ is the area of the plane $x_3 = 0$ covered by the rough  surface. As we are dealing with scattering from a randomly rough surface, it is the average of this function over the ensemble of realizations of the surface profile function that we have to calculate. The contribution to this average from the light scattered incoherently is 
\begin{align}
  \label{eq:10}
  \left \langle \frac{\partial R_{\alpha \beta}(\bqp|\bkp)}{\partial \Omega_s} \right\rangle_{\textrm{incoh}} 
  &= 
  \frac{1}{S}\left(\frac{\omega}{2 \pi c}\right)^2 \frac{\cos^2 \theta_s}{\cos \theta_0}
  \left[ \left< \left|R_{\alpha \beta}(\bqp|\bkp)\right|^2 \right> - \left| \Big< R_{\alpha \beta}(\bqp|\bkp)\Big> \right|^2 \right].
\end{align}

Closely related to the matrix of scattering amplitudes ${\bf R}(\bqp|\bkp)$ is the scattering matrix ${\bf S}(\bqp|\bkp)$ whose elements $\{ S_{\alpha \beta}(\bqp|\bkp)\}$ 
are given by 
\begin{align}
  \label{eq:11}
  S_{\alpha \beta}(\bqp|\bkp) & = \frac{\alpha_0^{1/2}(\qp)}{\alpha_0^{1/2}(\kp)} R_{\alpha \beta}(\bqp|\bkp).
\end{align}
These elements satisfy the reciprocity relations~\cite{6}
\begin{subequations}
\label{eq:12}
\begin{align}
  S_{pp}(\bqp|\bkp) & = S_{pp}(-\bkp|-\bqp) \label{eq:12a} \\
  S_{ss}(\bqp|\bkp) & = S_{ss}(-\bkp|-\bqp) \label{eq:12b} \\
  S_{ps}(\bqp|\bkp) & = -S_{sp}(-\bkp|-\bqp) \label{eq:12c}, 
\end{align}
\end{subequations}
which serve as a check on the correctness of their derivation. In terms of the elements of the scattering matrix, Eq.~\eqref{eq:9} takes the form
\begin{align}
  \label{eq:13}
 \left< \frac{\partial R_{\alpha \beta}(\bqp|\bkp)}{\partial \Omega_s} \right>_{\textrm{incoh}} 
  &  = 
  \frac{1}{S}\left(\frac{\omega}{2 \pi c}\right)^2 \cos \theta_s
  \left[ \left< \left| S_{\alpha \beta}(\bqp|\bkp) \right|^2 \right> - \left| \Big< S_{\alpha \beta}(\bqp|\bkp) \Big> \right|^2 \right].
\end{align}
This is the definition we will work with.

In the Appendix it is shown that the $ss$ element of the expression given by Eq.~\eqref{eq:13} obtained on the basis of second-order phase perturbation theory can be written as
\begin{align}
  \label{eq:14}
  \left\langle \frac{\partial R_{ss}(\bqp|\bkp)}{\partial \Omega_s} \right\rangle_{\textrm{incoh}}
  & = \frac{(\varepsilon - 1)^2}{(2 \pi)^2} \left(\frac{\omega}{c}\right)^6 
      \frac{\cos \theta_s}{\left[d_s(\qp)d_s(\kp)\right]^2}  
      \exp\left[-2 M(\bqp|\bkp)\right] 
  \nn \\ 
  & \quad \times 
      \sum\limits_{n = 1}^\infty \frac{ \left[4 \delta^2 \alpha_0(\qp) \alpha_0(\kp)
        (\pvecUnit{q} \cdot \pvecUnit{k})^2 \right]^n}{n !}  
      \int \dint[2]{\up} W^n(|\pvec{u}|) \exp\left[-\imu(\bqp - \bkp) \cdot \bup \right].
\end{align}
In writing this expression we have introduced the functions
\begin{subequations}
  \label{eq:15}
  \begin{align}
    d_p(\qp) & = \varepsilon \alpha_0(\qp) + \alpha(\qp) \label{eq:15a} \\
    d_s(\qp)& =  \alpha_0(\qp) + \alpha(\qp), \label{eq:15b}
  \end{align}
\end{subequations}
where  
\begin{align}
  \label{eq:def-alpha}
  \alpha(\qp) &= \left[\varepsilon \left(\frac{\w}{c}\right)^2 - \qp^2 \right]^{1/2} 
     \qquad \quad \Re\alpha(\qp) > 0, \,\, \Im\alpha(\qp) > 0. 
\end{align}
The function $ M(\bqp|\bkp)$ is given by [see the Appendix] 
\begin{align} 
  \label{eq:16} 
  M(\bqp|\bkp)  
  &= 
  2 \delta^2 \alpha_0^{1/2}(\qp) \alpha_0^{1/2}(\kp) 
  \int \dfint[2]{\pp}{(2\pi)^2} \Re F(\bqp|\bpp|\bkp) g(\left|\bpp - \bkp \right|), 
\end{align}
where 
\begin{align} 
  \label{eq:17}
  F(\bqp|\bpp|\bkp) 
  =&
  \sgn(\pvecUnit{q} \cdot \pvecUnit{k}) 
   \bigg\{ 
  \frac{1}{2}[\alpha(\qp) + \alpha(\kp)] (\pvecUnit{q} \cdot \pvecUnit{k})  
  + (\varepsilon - 1)   
    (\pvecUnit{q} \times \pvecUnit{p})_3          
   \frac{\alpha_0(\pp) \alpha(\pp)}{d_p(\pp)} 
   (\pvecUnit{p} \times \pvecUnit{k})_3   \nn \\  
  & \qquad \qquad  \qquad 
  -(\varepsilon - 1)\left(\frac{\w}{c} \right)^2
  \frac{ (\pvecUnit{q} \cdot \pvecUnit{p}) (\pvecUnit{p} \cdot \pvecUnit{k}) }{d_s(\pp)} 
   \bigg\}, 
\end{align}
with $\sgn(\cdot)$ denoting the sign function.
 
We now turn to an evaluation of the ingredients in Eq.~\eqref{eq:14}. We begin with the expression for $ 2M(\bqp|\bkp)$ given 
by Eqs.~\eqref{eq:16}--\eqref{eq:17}. With the use of Eqs.~\eqref{eq:5}--\eqref{eq:6} we rewrite it in terms of $W(\bxp )$:
\begin{align}  \label{eq:18}
  2 M(\bqp|\bkp)  
  & = 
  \delta^2 \alpha_0^{1/2}(\qp) \alpha_0^{1/2}(\kp) 
  \sgn(\pvecUnit{q} \cdot \pvecUnit{k}) 
   \bigg\{ 2[\alpha(\qp) + \alpha(\kp)]  (\pvecUnit{q} \cdot \pvecUnit{k}) \nn \\  
  &\quad 
    + \frac{(\varepsilon - 1)}{\pi^2}  
      \Re \int_0^\infty \dint{\pp} \pp \int_{-\pi}^{\pi} \dint{\phi_p}
      \int_0^\infty \dint{\xp} \xp W(\xp)\int_{-\pi}^{\pi} \dint{\phi_x} \exp[-\imu \pp\xp\cos(\phi_p - \phi_x)]  \nn \\
  &\qquad \times 
      \exp[\imu \kp\xp\cos(\phi_k - \phi_x)]
       \bigg[
             \frac{\alpha_0(\pp) \alpha(\pp)}{d_p(\pp)} \sin(\phi_p - \phi_q)\sin(\phi_k - \phi_p)  
              \nn \\ & \qquad  \qquad  \qquad \qquad  \qquad  \qquad  \qquad 
              - \frac{(\w/c)^2}{d_s(\pp)}\cos(\phi_q - \phi_p)\cos(\phi_p - \phi_k) 
       \bigg] \bigg\},
\end{align}
where $\phi_q$, $\phi_p$, $\phi_k$, and $\phi_x$ are the azimuthal angles of the unit vectors 
$\pvecUnit{q}$, $\pvecUnit{p}$, $\pvecUnit{k}$, and $\pvecUnit{x}$, respectively, 
measured from the positive $x_1$ axis [see Fig.~\ref{fig:geometry}]. On evaluating the angular integrals this result becomes
\begin{align}     
  \label{eq:19}
  2 M(\bqp|\bkp) & = 2\delta^2 \alpha_0^{1/2}(\qp) \alpha_0^{1/2}(\kp)
  \left| \pvecUnit{q} \cdot \pvecUnit{k} \right| 
  \bigg\{ \alpha(\qp) + \alpha(\kp)   \nn \\
  &\quad - (\varepsilon - 1)\, \Re 
   \bigg[
     \int_0^\infty \dint{\pp} \pp 
     \left( 
       \frac{\alpha_0(\pp) \alpha(\pp)}{d_p(\pp)} 
       +
       \frac{(\w/c)^2}{d_s(\pp)}
     \right)  
     \int_0^\infty \dint{\xp} \xp W(\xp) J_0(\pp \xp)J_0(\kp \xp) 
  \nn \\
  & \qquad \qquad \qquad \quad 
  +
   \int_0^\infty \dint{\pp} \pp 
     \left( 
       -\frac{\alpha_0(\pp) \alpha(\pp)}{d_p(\pp)} 
       +
       \frac{(\w/c)^2}{d_s(\pp)}
     \right)  
     \int_0^\infty \dint{\xp} \xp W(\xp) J_2(\pp \xp)J_2(\kp \xp)
     \bigg]\bigg\}, 
\end{align}
where $J_n(z)$ is a Bessel function of the first kind and order $n$, and we have used the relation $x\sgn(x)=|x|$. Finally, due to the circular symmetry of $W(|\pvec{u}|)$ we obtain the result
\begin{align}
  \label{eq:22}
  \int \dint[2]{\up} W^n(|\pvec{u}|) \mathrm{exp}[-i(\bqp - \bkp) \cdot \bup] 
  &= 2\pi \int_0^\infty \dint{\up} \up W^n(\up) J_0(\left| \bqp - \bkp \right| \up).
\end{align}

In the case of normal incidence ($\pvec{k} = \vec{0}$) and in-plane ($\pvecUnit{q}\parallel \pvecUnit{k}$) scattering, Eq.~\eqref{eq:14} becomes
\begin{align} 
  \label{eq:24}
  \left \langle \frac{\partial R_{ss}(\bqp|\bzero)}{\partial \Omega_s} \right\rangle_{\textrm{incoh}}  &\equiv \left \langle \frac{\partial R_{ss}(\theta_s)}{\partial \Omega_s} \right\rangle_{\textrm{incoh}}  \nn\\
&= \frac{(\varepsilon - 1)^2}{(2 \pi)^2} \left(\frac{\omega}{c}\right)^6 
\frac{\cos \theta_s}{\left[d_s(\qp)d_s(0)\right]^2}  
\exp\left[-2 M(\bqp|\bzero)\right] \nn\\
&\quad \times 
\sum\limits_{n = 1}^\infty \frac{[4 \delta^2 \alpha_0(\qp) \alpha_0(0)]^n}{n !} 
 \int \dint[2]{\up} W^n(|\pvec{u}|) \mathrm{exp}(-\imu \bqp \cdot \bup),
\end{align}
and Eq.~\eqref{eq:19} can be written as
\begin{align}
  \label{eq:25}
  2 M(\bqp|\bzero) 
    & = 2\delta^2 \alpha_0^{1/2}(\qp) \alpha_0^{1/2}(0) 
          \bigg\{\alpha(\qp) + \alpha(0)  \nn \\
    & \quad -(\varepsilon -1) \, \Re \int_0^\infty \dint{\pp} \pp\left[\frac{\alpha_0(\pp) \alpha(\pp)}{d_p(\pp)} + \frac{(\w/c)^2}{d_s(\pp)}\right] 
    \int_0^\infty \dint{\xp} \xp W(\xp)J_0(\pp \xp) \bigg\}. 
\end{align}
From Eq.~\eqref{eq:9-1} it is noted that for normal incidence $\pvecUnit{k}=(\cos\phi_0,\sin\phi_0,0)$, even if $\pvec{k}/k_\parallel$ is not well-defined in this case. Moreover, for scattering into directions that are normal to the mean surface we have $\pvecUnit{q}=\pvecUnit{k}$, so for all directions $\pvec{q}$ corresponding to in-plane scattering $|\pvecUnit{q} \cdot \pvecUnit{k}|=1$. These results were used in arriving at the expression presented in  Eqs.~\eqref{eq:24} and \eqref{eq:25}.

\section{The Inverse Problem}
\label{Sec:Inversion}
 
 To determine the function $W(\xp )$ from scattering data for $\la \p R_{ss}(\theta_s)/\p\Omega_s\ra_{\textrm{incoh,input}}$, we assume an analytic form for it that contains adjustable parameters.  The values of these parameters, together with the rms height $\delta$, are determined by minimizing a cost function with respect to variations of these parameters.  The cost function we use is
 \begin{align}
 \chi^2({\cal P}) = \int\limits^{\frac{\pi}{2}}_{-\frac{\pi}{2}} d\theta_s \bigg[\bigg\la \frac{\p R_{ss}(\theta_s)}{\p\Omega_s}\bigg\ra_{\textrm{incoh,input}}
  - \left\la \frac{\p R_{ss}(\theta_s)}{\p\Omega_s}\right\ra_{\textrm{incoh,calc}}\bigg]^2 , 
\label{eq:27}
 \end{align}
where ${\cal P}$ denotes the set of variational parameters used to characterize $\la \p R_{ss}(\theta_s)/\p\Omega_s\ra_{\textrm{incoh,calc}}$.  The minimization of this function with respect to the elements of ${\cal P}$ was carried out by the use of the routine ``\textrm{lmdif1}'' contained in the Fortran package MINPACK which is part of  the general purpose mathematical library SLATEC~\cite{SLATEC}. The routine \textrm{lmdif1} implements a modified version of the Levenberg-Marquardt algorithm~\cite{LM1,LM2}, and it calculates the Jacobian by a forward-difference approximation.

The function $\la \p R_{ss}(\theta_s)/\p \Omega_s\ra_{\textrm{incoh,input}}$ was obtained from rigorous, nonperturbative, purely numerical solutions~\cite{6,7} of the reduced Rayleigh equation for the scattering of polarized light from a two-dimensional randomly rough penetrable dielectric surface~\cite{8}.  These calculations were carried out for an ensemble of random surfaces generated~\cite{7} on the basis of expressions for $W(|\bxp |)$ of either the  exponential form
\begin{subequations}
  \label{eq:28}
  \begin{align}
    W(|\pvec{x} |) = \exp \left( - \frac{x_\parallel}{a}\right), 
    \label{eq:28_exponential}
  \end{align}
  or the Gaussian form
  \begin{align}
    W(|\pvec{x} |) = \exp \left\{- \left( \frac{x_\parallel}{a} \right)^2 \right\}. 
    \label{eq:28_Gaussian}
  \end{align}
\end{subequations}
In Eqs.~\eqref{eq:28}, $a$ denotes the transverse correlation length of the surface roughness. 

The function $\la \p R_{ss}(\theta_s)/\p \Omega_s\ra_{\textrm{incoh,calc}}$ was obtained by evaluating the expression for it obtained using phase perturbation theory~[Eq.~\eqref{eq:14}] for the trial function assumed to represent $W(|\pvec{x} |)$.  Several forms for this trial function were used in our calculations.  In the first set of forms we assumed an exponential or Gaussian trial function, that is,
\begin{subequations}
  \label{eq:29}
  \begin{align}
    W(|\pvec{x} |) = \exp \left( - \frac{x_\parallel}{a^\star}\right), 
    \label{eq:29_exponential}
  \end{align}
  or
  \begin{align}
    W(|\pvec{x} |) = \exp \left\{- \left( \frac{x_\parallel}{a^\star} \right)^2 \right\}. 
    \label{eq:29_Gaussian}
  \end{align}
\end{subequations}
In this case the variational parameters of the reconstruction are $\delta^\star, a^\star$, and potentially also $\e^\star$.  For the  second set of forms for the trial function a stretched exponential was assumed 
\begin{align}
  W(|\bxp |) = \exp \left\{-\left(\frac{\xp}{a^\star}\right)^{\gamma^\star} \right\}, 
  \label{eq:30}
\end{align}
which reduces to the exponential and Gaussian forms when $\gamma^\star=1$ and $\gamma^\star=2$, respectively.
In this case the variational parameters of the reconstruction are $\delta^\star, a^\star, \gamma^\star$, and potentially $\e^\star$.

\section{Results}
\label{sec:Results}

We will now illustrate the inversion method developed here by applying it to the reconstruction of $W(|\bxp |)$, first by the use of one of the trial functions~\eqref{eq:29} and then by the use of the more general trial function (\ref{eq:30}).

\subsection{Exponentially correlated surface roughness}
\label{sec:Result-Exponential}

%
\begin{figure}[tbp] 
  \centering
  \includegraphics*[width=0.45 \columnwidth]{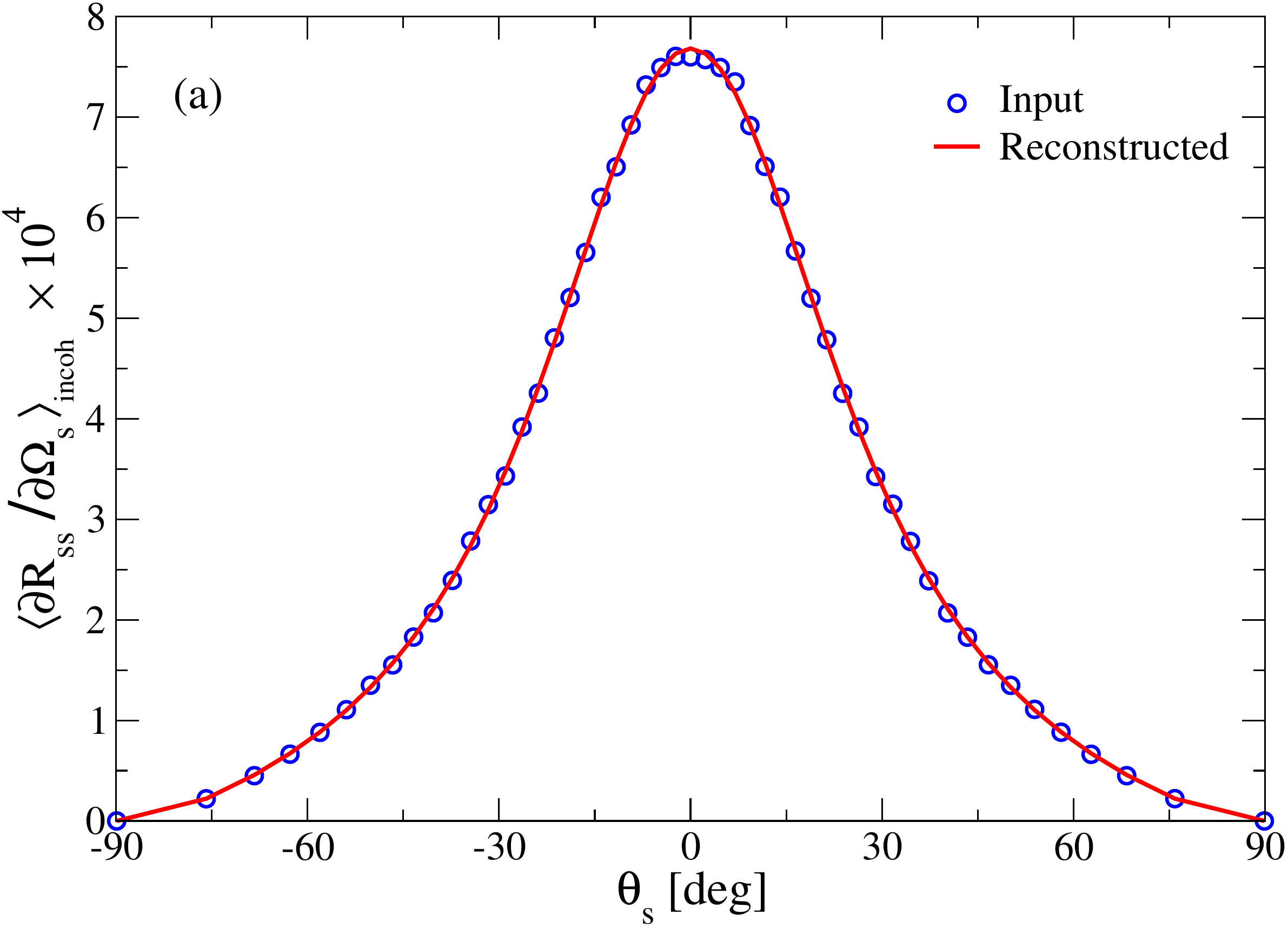}\qquad
  \includegraphics*[width=0.45 \columnwidth]{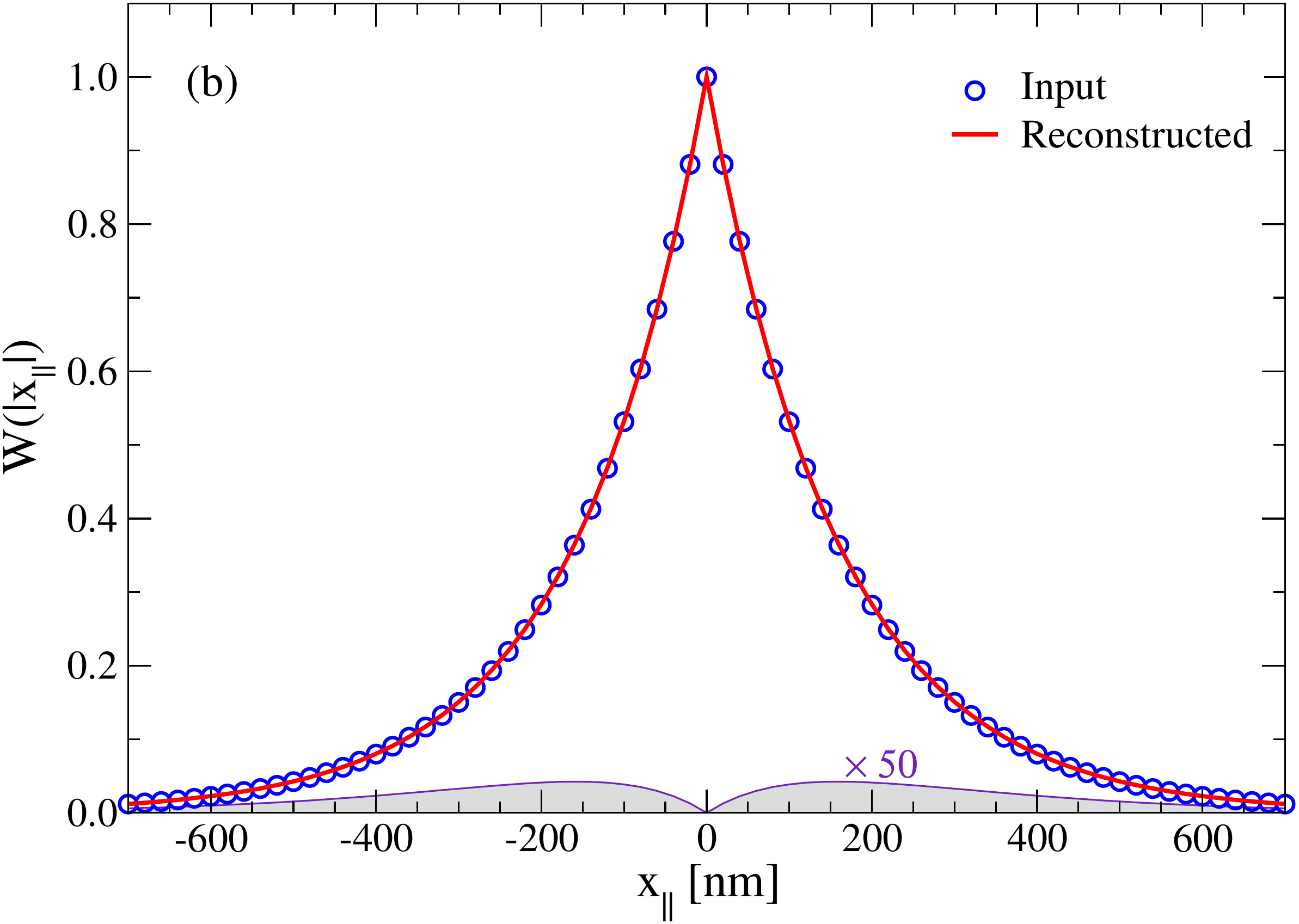}
  \caption{Reconstruction of the rms-roughness $\delta^\star$ and transverse correlation length $a^\star$ from in-plane scattering data obtained for exponentially correlated surfaces.  (a) The incoherent component of the in-plane, co-polarized (s-to-s) mean differential reflection coefficient $\la \p R_{ss}/\p\Omega_s\ra_{\textrm{incoh}}$ as a function of the polar angle of scattering $\theta_s$ obtained from computer simulations~(open circles), and from second-order phase perturbation theory with the use of the reconstructed surface roughness parameters (solid curve), for a two-dimensional randomly rough dielectric surface defined by Eq.~\protect\eqref{eq:29_exponential}.  The surface roughness parameters assumed in the computer simulations have the values $\delta =\SI{9.50}{\nano\meter}$ and $a = \SI{158.20}{\nano\meter}$, while the reconstructed values of these parameters are $\delta^\star = \SI{9.519}{\nano\metre}$, and $a^\star = \SI{158.565}{\nano\metre}$.  The dielectric constant of the substrate is $\e = 2.64$, and the wavelength of the s-polarized light incident normally on the mean surface is $\lambda = \SI{632.8}{\nano\meter}$.  (b) The input (open circles) and reconstructed (solid curve) surface height autocorrelation function $W(|\bxp |)$ for the random surface. The shaded gray region represents the absolute difference between the input and reconstructed surface height autocorrelation functions.}
%
  \label{fig:1}
\end{figure}
%

\begin{figure}[tbph] 
  \centering
  \includegraphics*[width=0.45 \columnwidth]{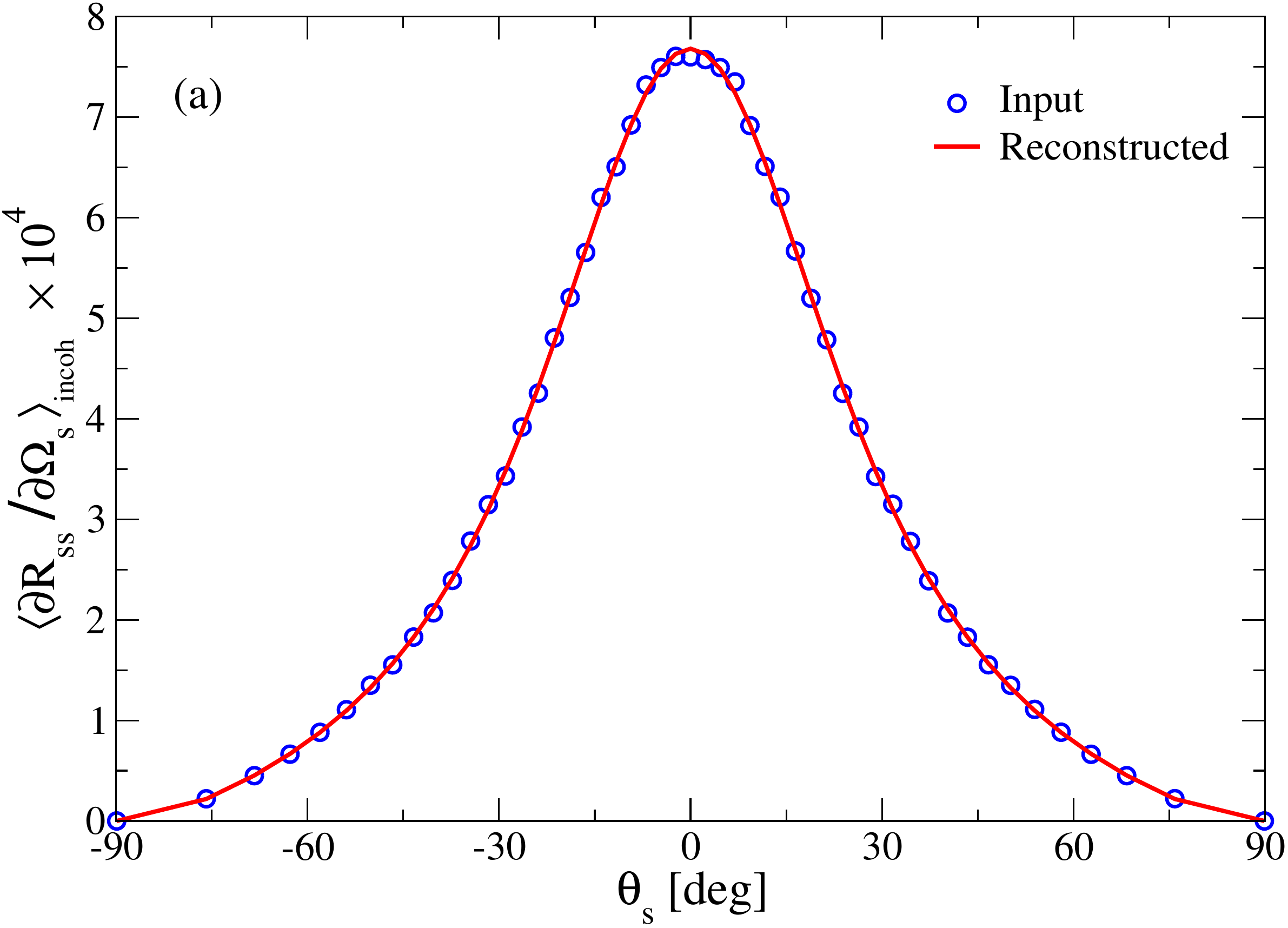}\qquad
  \includegraphics*[width=0.45 \columnwidth]{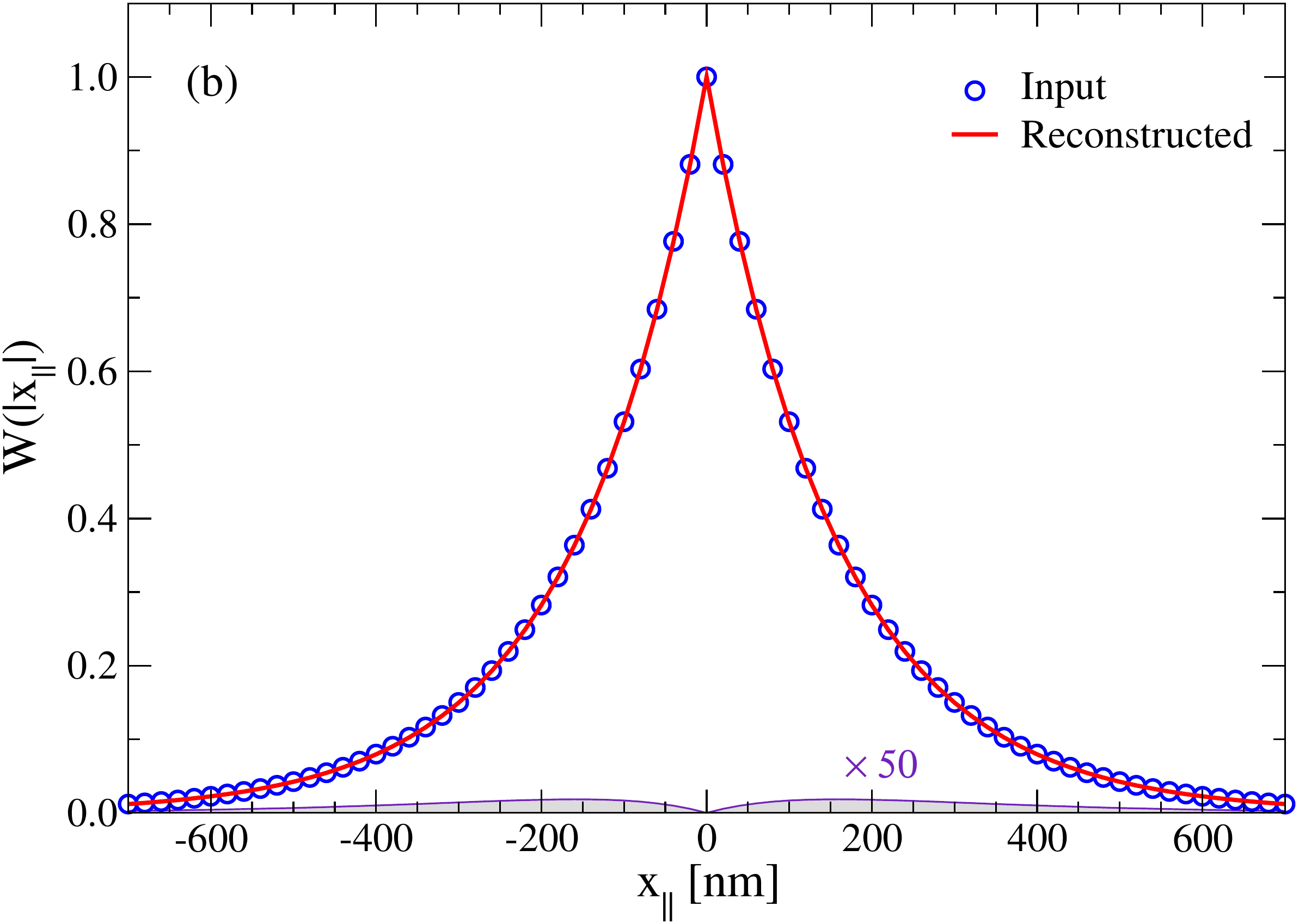}
  \caption{Reconstruction of the rms-roughness $\delta^\star$, the transverse correlation length $a^\star,$ and the dielectric constant of the substrate $\e^\star$ from the in-plane scattering data.  This figure is the same as Fig.~\protect\ref{fig:1} except now the dielectric constant of the substrate is also reconstructed.  The reconstructed surface roughness parameters are found to be $\delta^\star = \SI{9.272}{\nano\metre}$, $a^\star =\SI{158.042}{\nano\metre}$, and the reconstructed dielectric constant has the value $\e^\star = 2.718$.}
  \label{fig:2}
\end{figure}

\begin{figure}[btph] 
  \centering
  \includegraphics*[width=0.45 \columnwidth]{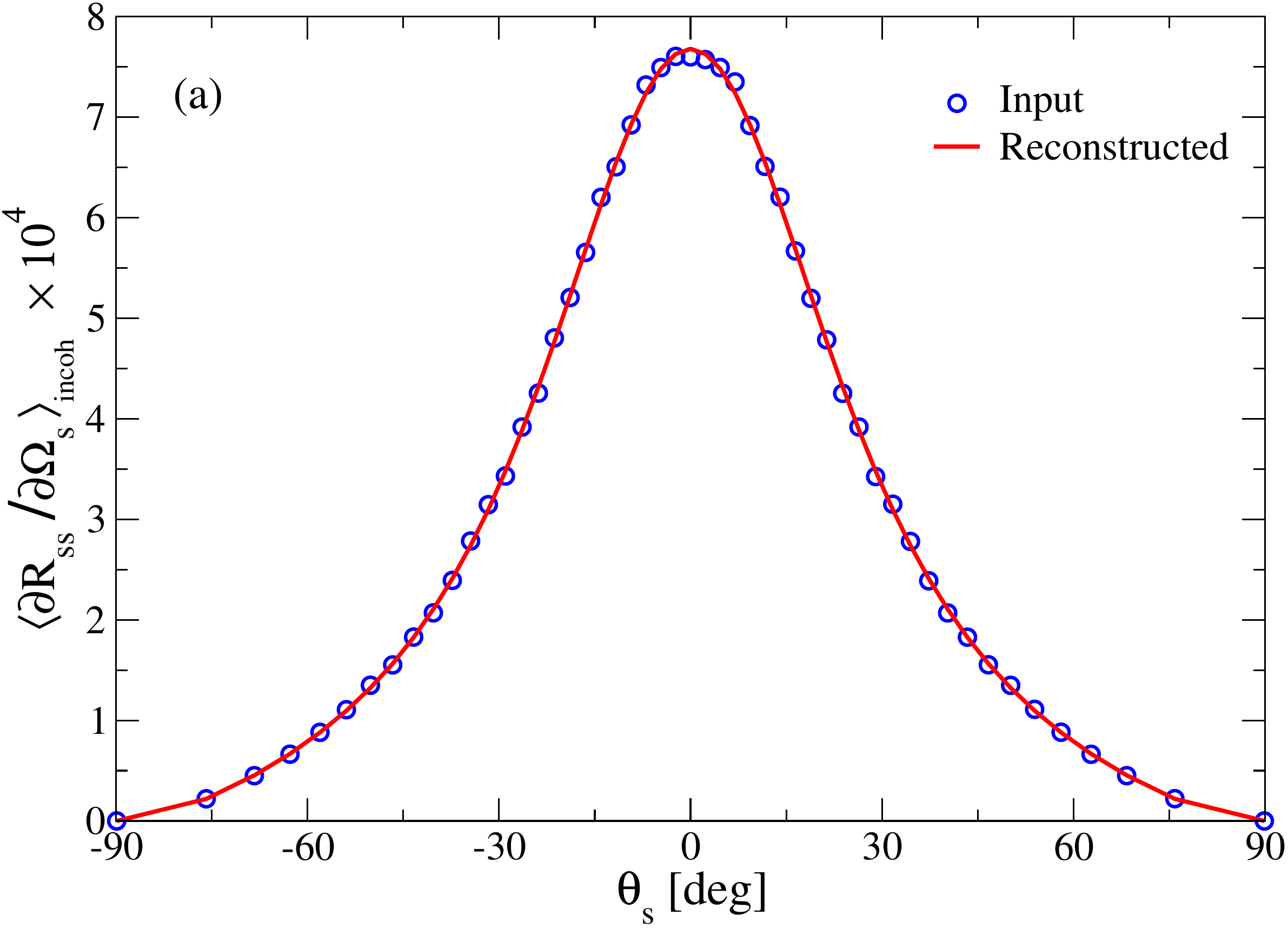}\qquad
  \includegraphics*[width=0.45 \columnwidth]{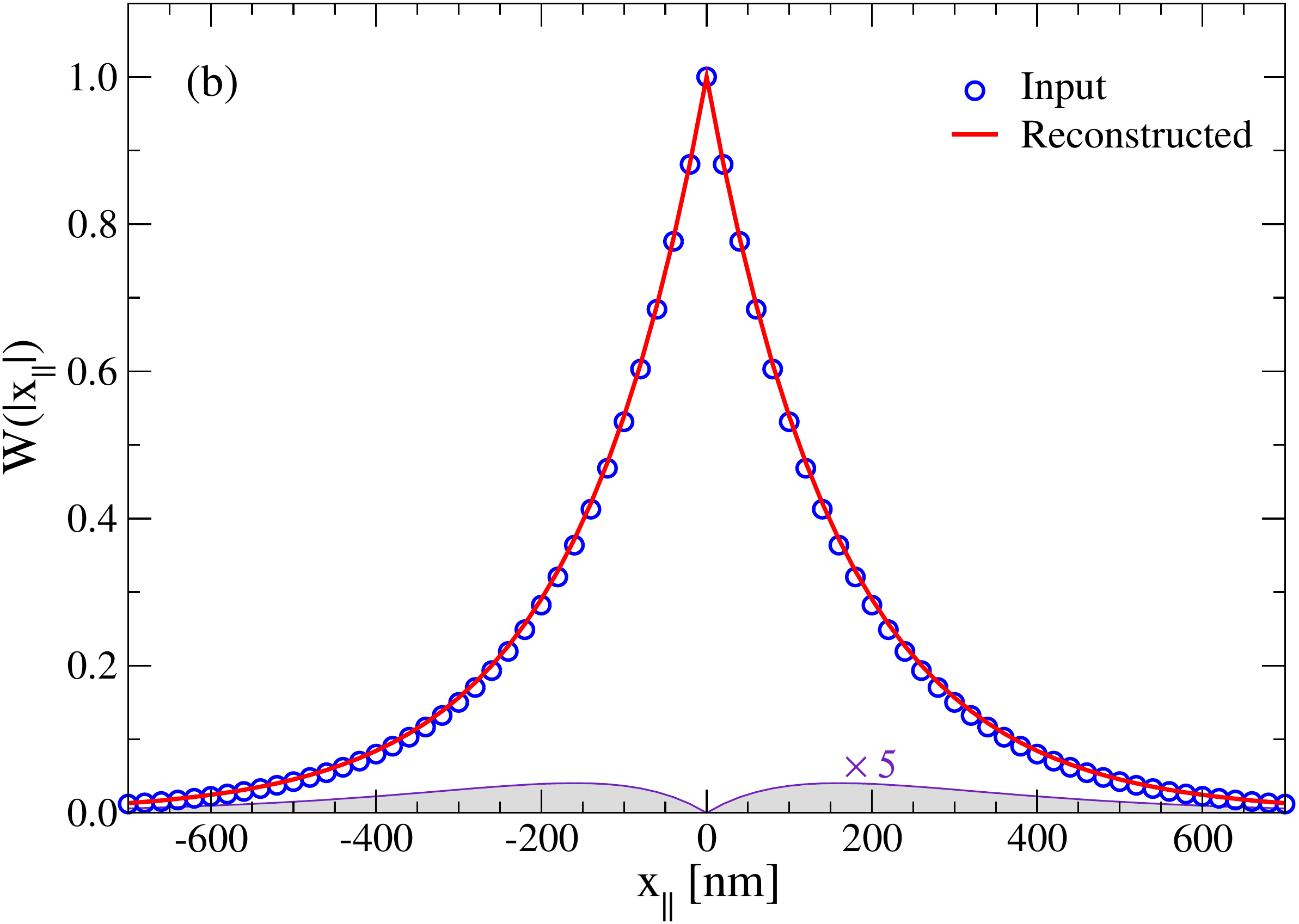}
  \caption{Reconstruction of the rms-roughness $\delta^\star$, the transverse correlation length $a^\star$, and the exponent $\gamma^\star$ from in-plane scattering data.  This figure is the same as Fig.~\protect\ref{fig:1} except that now the  trial function for $W(|\bxp |)$ has the stretched  exponential form given by Eq.~\protect\eqref{eq:30}.  The reconstructed surface roughness parameters are found to have the values $\delta^\star=\SI{9.425}{\nano\metre}$, $a^\star =\SI{161.717}{\nano\metre}$, and $\gamma^\star = 1.012$.}
  \label{fig:3}
\end{figure}

\begin{figure}[tbph] 
  \centering
\includegraphics*[width=0.45 \columnwidth]{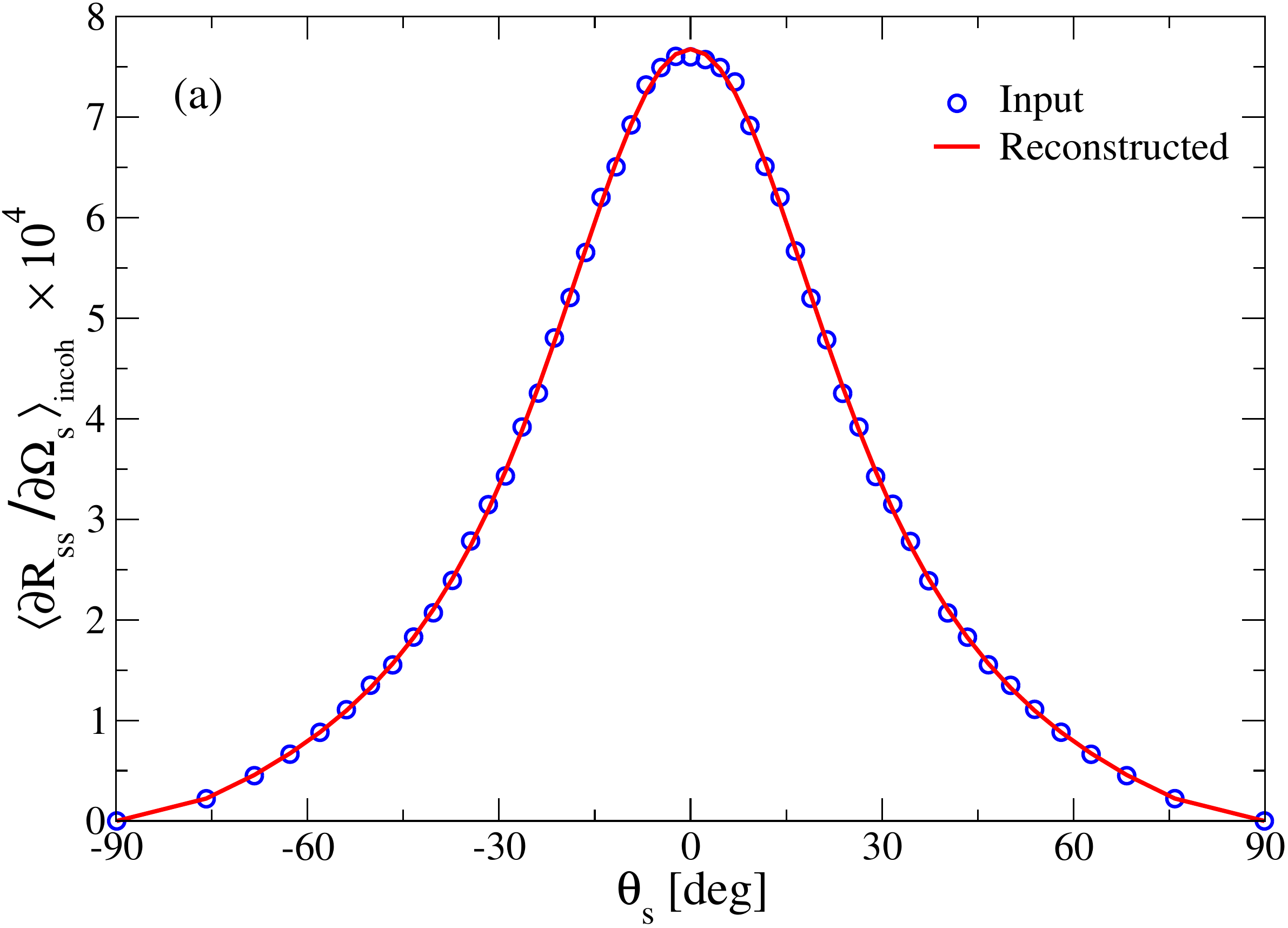}\qquad
\includegraphics*[width=0.45 \columnwidth]{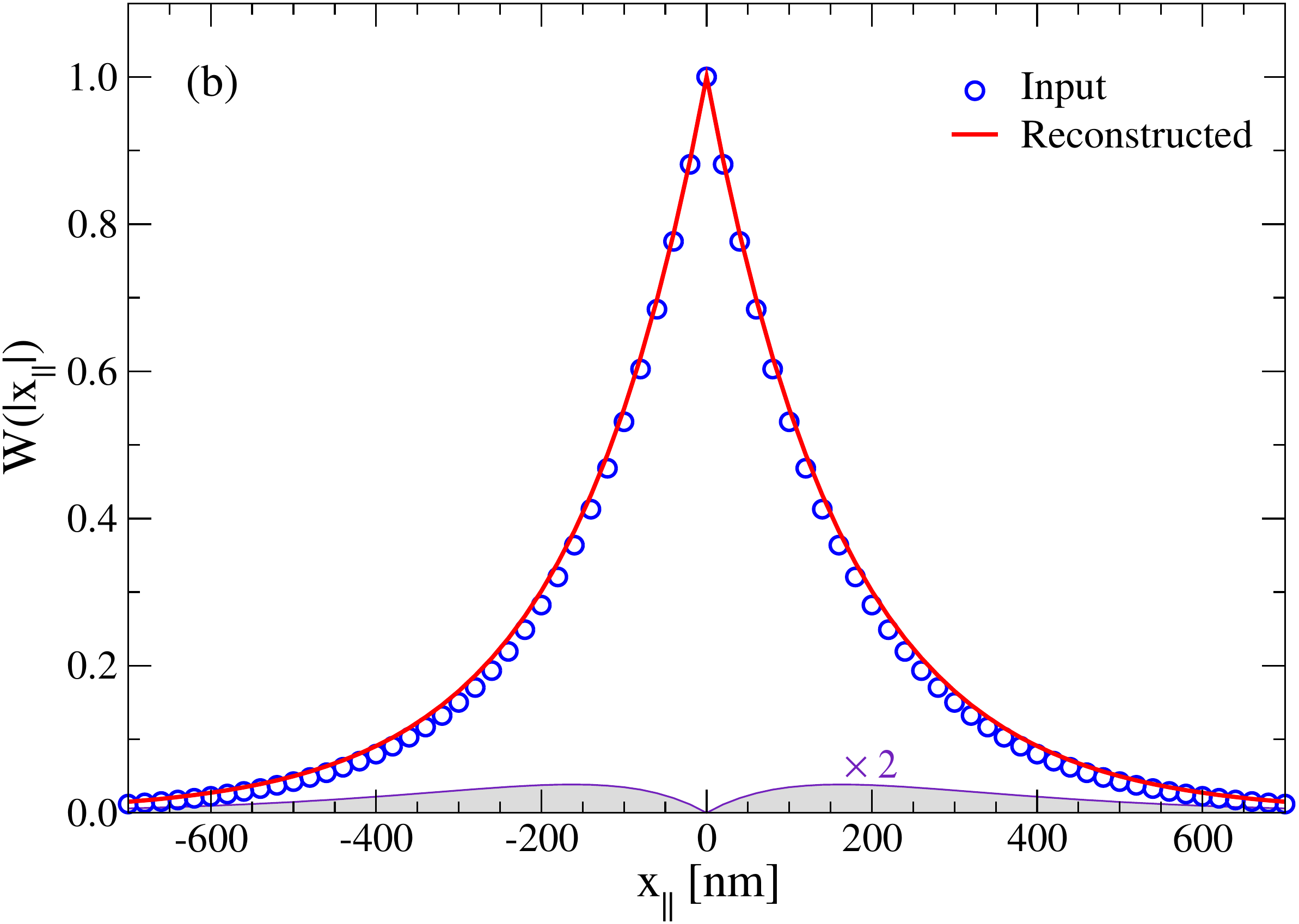}
\caption{Reconstruction of the rms-roughness $\delta^\star$, the transverse correlation length $a^\star$, the exponent $\gamma^\star$, and the dielectric constant of the substrate $\e^\star$ from in-plane scattering data.  This figure is the same as Fig.~\protect\ref{fig:1} except that now the trial function for $W (|\bxp |)$ has the stretched exponential form given by Eq.~\eqref{eq:30}, and the dielectric constant of the scattering medium is assumed to be unknown.  The reconstructed surface roughness parameters are found to have the values $\delta^\star = \SI{9.774}{\nano\metre}$, $a^\star =\SI{166.709}{\nano\metre}$, $\gamma^\star = 1.027$, and the reconstructed value of the dielectric constant is $\e^\star = 2.507$.}
\label{fig:4}
\end{figure}

For the first scattering system we consider, it is assumed that the surface height autocorrelation function $W(|\pvec{x}|)$ is \textit{exponential}, Eq.~\eqref{eq:28_exponential}, and characterized by a transverse correlation length $a = \SI{158.20}{\nano\meter}$ and an  rms height of the surface  $\delta = \SI{9.50}{\nano\meter}$. The medium above the surface is vacuum and the dielectric constant of the substrate is $\e = 2.64$ (photoresist). 
The wavelength (in vacuum) of the s-polarized incident light is $\lambda = \SI{632.8}{\nano\meter}$. 
%
For this geometry and by the method of Ref.~\cite{7}, we calculated the mean differential reflection coefficients by averaging the results from $\num{5 000}$ realizations of the surface profile function. For normal incidence, the in-plane, s-to-s co-polarized incoherent component of the mean differential reflection coefficient~(DRC) obtained in this way is presented as a function of the scattering angle $\theta_s$ by open circles in Fig.~\ref{fig:1}(a) [the same data set also appears in Figs.~\ref{fig:2}(a)--~\ref{fig:4}(a)]. These data constitute the input function $\la R_{ss}(\theta_s)/\p\Omega_s\ra_{\textrm{incoh,input}}$ for our first set of reconstruction examples.


As our first example of reconstruction based on this data set, we assume that the trial function $W(|\bxp |)$ has the exponential form given by Eq.~\eqref{eq:29_exponential}. The set of variational parameters is therefore ${\cal P} = \{ \delta^\star, a^\star\}$.  The use of a mean differential reflection coefficient generated by the use of a known $W(|\pvec{x}|)$ in our inversion approach enables us to assess the quality of the reconstructions we obtain. By starting the minimization procedure with the values $\delta^\star = \SI{2.00}{\nano\meter}$ and $a^\star = \SI{75.00}{\nano\meter}$, the values of these parameters that minimize the cost function $\chi^2({\cal P})$, Eq.~\eqref{eq:27}, were found to be $\delta^\star =\SI{9.519}{\nano\metre}$ and $a^\star =\SI{158.565}{\nano\meter}$, to be compared with the values $\delta = \SI{9.50}{\nano\meter}$ and $a = \SI{158.20}{\nano\meter}$ used to generate  the input data.  In the minimization procedure we assumed that both $\delta^\star$ and $a^\star$ were restricted to positive values.  The inversion is quite accurate.  The function $\la \p R_{ss}(\theta_s)/\p\Omega_s\ra_{\textrm{incoh,calc}}$ calculated  with the reconstructed values of $\delta^\star$ and $a^\star$ by means of second-order phase perturbation theory is plotted as the solid curve in Fig.~\ref{fig:1}(a) while the reconstructed correlation function $W(|\pvec{x}|)$ is plotted as the solid curve in Fig.~\ref{fig:1}(b).  The reconstructed $W(|\pvec{x}|)$ is nearly superimposed on the input $W(|\pvec{x}|)$ [open symbols Fig.~\ref{fig:1}(b)]. The shaded region in Fig.~\ref{fig:1}(b), and in the subsequent plots of $W(|\pvec{x}|)$, represents the magnitude of the difference between the input and reconstructed values of this function.  This difference is seen to be very small.

%
%

In the preceding example it was assumed that the dielectric constant of the scattering medium was known.  For our second  example we take the input data from our first example, given by the open circles in Fig.~\ref{fig:1}(a), but now assume that together with the roughness parameters the dielectric constant of the substrate is unknown.  Therefore the variational parameter set is now ${\cal P} = \{ \delta^\star, a^\star, \e^\star\}$.  The results of this inversion are shown in Fig.~\ref{fig:2}, and it is seen that also in this case a rather good reconstruction is  obtained.  By starting the minimization procedure with the values $\delta^\star = \SI{2.00}{\nano\meter}$, $a^\star = \SI{75.00}{\nano\meter}$, and $\e^\star = 2.00$, the values of these parameters that minimize the cost function $\chi^2({\cal P})$ were determined to be $\delta^\star = \SI{9.272}{\nano\metre}$, $a^\star =\SI{158.042}{\nano\metre}$, and $\e^\star = 2.718$.  These are to be compared with the input values  $\delta = \SI{9.50}{\nano\meter}$, $a = \SI{158.20}{\nano\meter}$, and $\e = 2.64$. The function $\la \p R_{ss}(\theta_s /\p\Omega_s\ra_{\textrm{incoh,calc}}$ calculated with the reconstructed values of $\delta^\star, a^\star, \e^\star$ by means of second-order phase perturbation theory is plotted as the solid curve in Fig.~\ref{fig:2}(a), while the reconstructed correlation function $W(|\bxp |)$ is plotted as the solid curve in Fig.~\ref{fig:2}(b).  From a comparison of the results presented  in Figs.~\ref{fig:1} and \ref{fig:2} it is seen that the addition of a single variational parameter changes the reconstruction of $W(|\bxp |)$ only marginally.

%
%
A more stringent test of our inversion scheme is obtained when the trial function for $W(|\bxp |)$ has a functional form that differs from the form assumed in generating the input data that the reconstruction is based on. As our third example,  we therefore present results of our calculations when the trial  $W(|\bxp |)$ is assumed to have the stretched exponential form given by Eq.~\eqref{eq:30}.  The set of variational parameters is now ${\cal P} = \{ \delta^\star, a^\star, \gamma^\star \}$.  By starting the minimization procedure with the values $\delta^\star = \SI{2.00}{\nano\meter}$, $a^\star = \SI{75.00}{\nano\meter}$, and $\gamma^\star = 2.00$, the values of these parameters that minimize the cost function were found to be $\delta^\star = \SI{9.425}{\nano\metre}$, $a^\star =\SI{161.717}{\nano\metre}$, and $\gamma^\star = 1.012$. These values are fairly close to the input values $\delta = \SI{9.50}{\nano\meter}$, $a = \SI{158.20}{\nano\meter}$, and $\gamma = 1.0$ used in obtaining the simulation data.  However, the importance of this example is to show that our minimization procedure is in fact able to distinguish a Gaussian form for the correlation function from an exponential form. In Fig.~\ref{fig:3}(a) we plot the function $\la \p R_{ss}(\theta_s)/\p\Omega_s\ra_{\textrm{incoh,calc}}$ calculated by means of the second-order phase perturbation theory for the reconstructed values of $\delta^\star, a^\star, \gamma^\star$ (solid curve), together with a plot of the input function (open circles).  The agreement between these two results is quite good.  In Fig.~\ref{fig:3}(b) we present plots of the input (open circles) and reconstructed (solid curve) correlation functions $W(|\bxp |)$.  The latter curve very nearly coincides with the former curve.
%

%
%
%
In our final example assuming an exponentially correlated surface, we again use the stretched  exponential trial function for $W(|\bxp |)$, but now also assume that the dielectric constant of the scattering medium is unknown.  The set of variational parameters is now ${\cal P} = \{ \delta^\star, a^\star,\gamma^\star , \e^\star\}$.  We start the minimization of the cost function $\chi^2({\cal P})$ with the values $\delta^\star = \SI{2.00}{\nano\meter}$, $a^\star = \SI{75.00}{\nano\meter}$, and $\gamma^\star = 2.00$, and $\e^\star = 2.00$.
The values of these parameters that minimize the cost function are found to be $\delta^\star = \SI{9.774}{\nano\metre}$, $a^\star =\SI{166.709}{\nano\metre}$, $\gamma^\star = 1.027$, and $\e^\star = 2.507$. The proximity of these values to the input values, that are the same as those used previously, is poorer than for the first three examples, but the reconstructed values are still quite satisfactory.  The reconstructed function $\la \p R_{ss}(\theta_s )/\p\Omega_s\ra_{\textrm{incoh,calc}}$ [Fig.~\ref{fig:4}(a)] and the reconstructed correlation function $W(|\bxp |)$ [Fig.~\ref{fig:4}(b)],  calculated with these reconstructed values are still in good quantitative agreement with the corresponding input functions.

%
\begin{table}
  \caption{\label{tab:Exponential}  Summary of the scattering system parameters obtained during the different  reconstruction scenarios based on in-plane, s-to-s co-polarized scattering data corresponding to an \textit{exponentially} correlated surface; $\delta^\star$, $a^\star$, $\e^\star$, and $\gamma^\star$.  The scattering system parameters assumed in generating the input data were:  $\delta=\SI{9.50}{nm}$, $a=\SI{158.20}{nm}$, $\varepsilon=\num{2.64}$, and  $\theta_0=\ang{0}$. Note that an exponential correlation function corresponds to the exponent $\gamma=\num{1}$ for the stretched exponential. The last column indicates the relevant figure where the results of the reconstruction in question is presented. The symbol ``---'' indicates that the corresponding variable was not reconstructed and instead had the value assumed in the input data (numerical simulations). In the two first reconstructions a trial correlation function of the form~\eqref{eq:29_exponential} was used, while in the last two the form~\eqref{eq:30} was assumed.  }
  \begin{ruledtabular}
    \begin{tabular}{lllll}
      $\delta^\star \; \si{[nm]}$ & $a^\star \; \si{[nm]}$ & $\varepsilon^\star$ & $\gamma^\star$ & Comments \quad \\
      \hline 
      \num{9.519}         & \num{158.565}        & ---                & ---          &  Fig.~\protect\ref{fig:1}  \\
      \num{9.272}         & \num{158.042}        & \num{2.718}        & ---          &  Fig.~\protect\ref{fig:2}  \\
      \num{9.425}         & \num{161.717}        & ---                & \num{1.012}  &  Fig.~\protect\ref{fig:3}  \\
      \num{9.774}         & \num{166.709}        & \num{2.507}        & \num{1.027}  &  Fig.~\protect\ref{fig:4}  \\
    \end{tabular}
  \end{ruledtabular}
\end{table}
%
The parameters for the scattering system obtained in the different reconstruction scenarios detailed in this subsection are summarized in Table~\ref{tab:Exponential}.

\subsection{Gaussianly correlated surface roughness}
\label{sec:Result-Gaussian}
%
The second scattering system, from which we will use data for the purpose of inversion, is characterized by a \textit{Gaussian} correlation function instead of the exponential correlation function assumed in the earlier set of examples. Compared to the previous scattering system, in addition to the different form of $W(|\pvec{x}|)$, the only parameters that have changed are the rms-roughness and the dielectric constant of the substrate; they now take the values  $\delta=\SI{15.82}{\nano\meter}$ (an increase of more than $65\%$ compared to its previous value), and $\e=2.6896$, respectively. Except for the angles of incidence, all other parameters characterizing the scattering system remained unchanged, i.e., $a=\SI{158.20}{\nano\meter}$ and $\lambda = \SI{632.8}{\nano\meter}$. 

\begin{figure}[tbph] 
  \centering
  \includegraphics*[height=0.32\columnwidth]{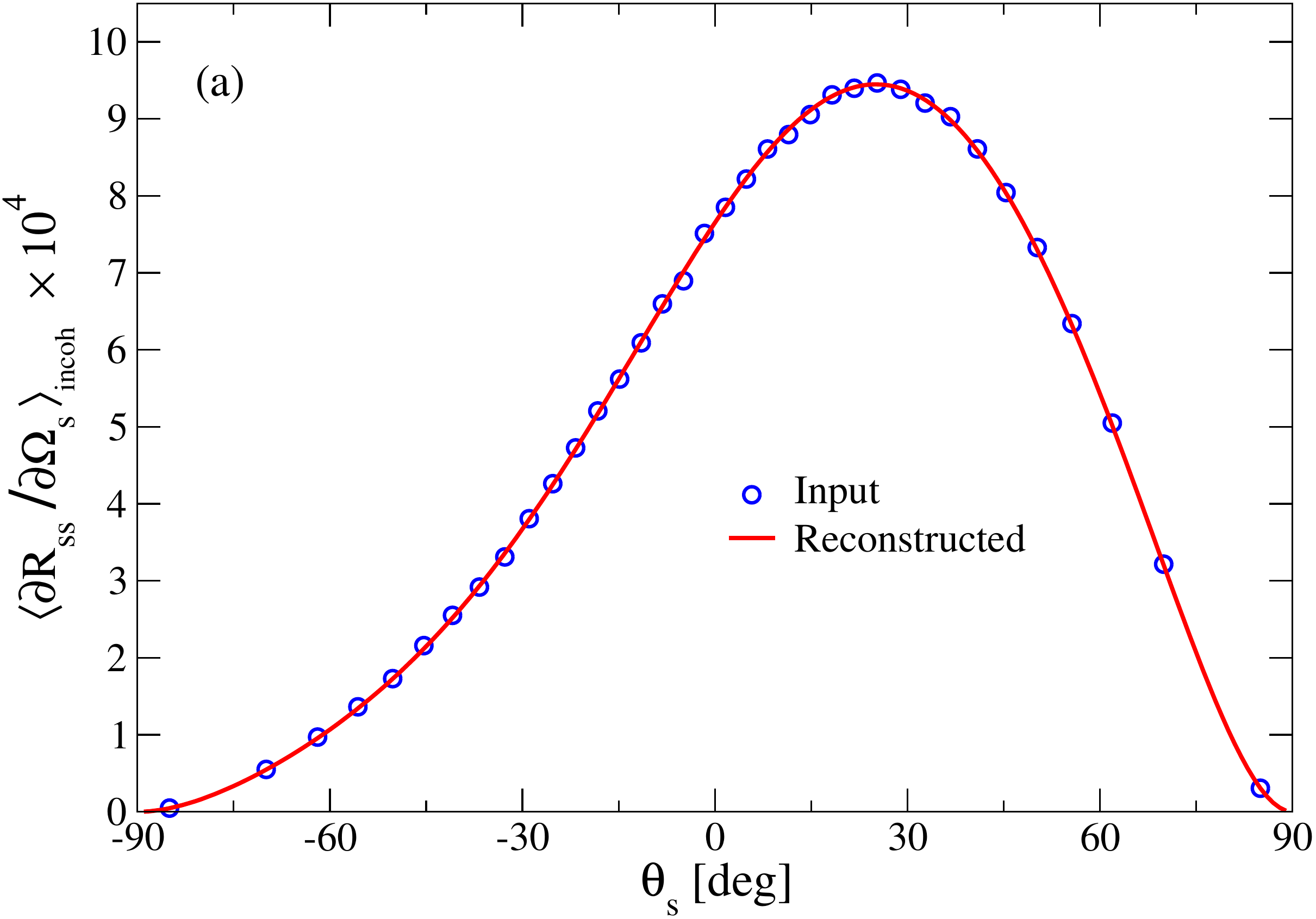} \qquad
  \includegraphics*[height=0.32\columnwidth]{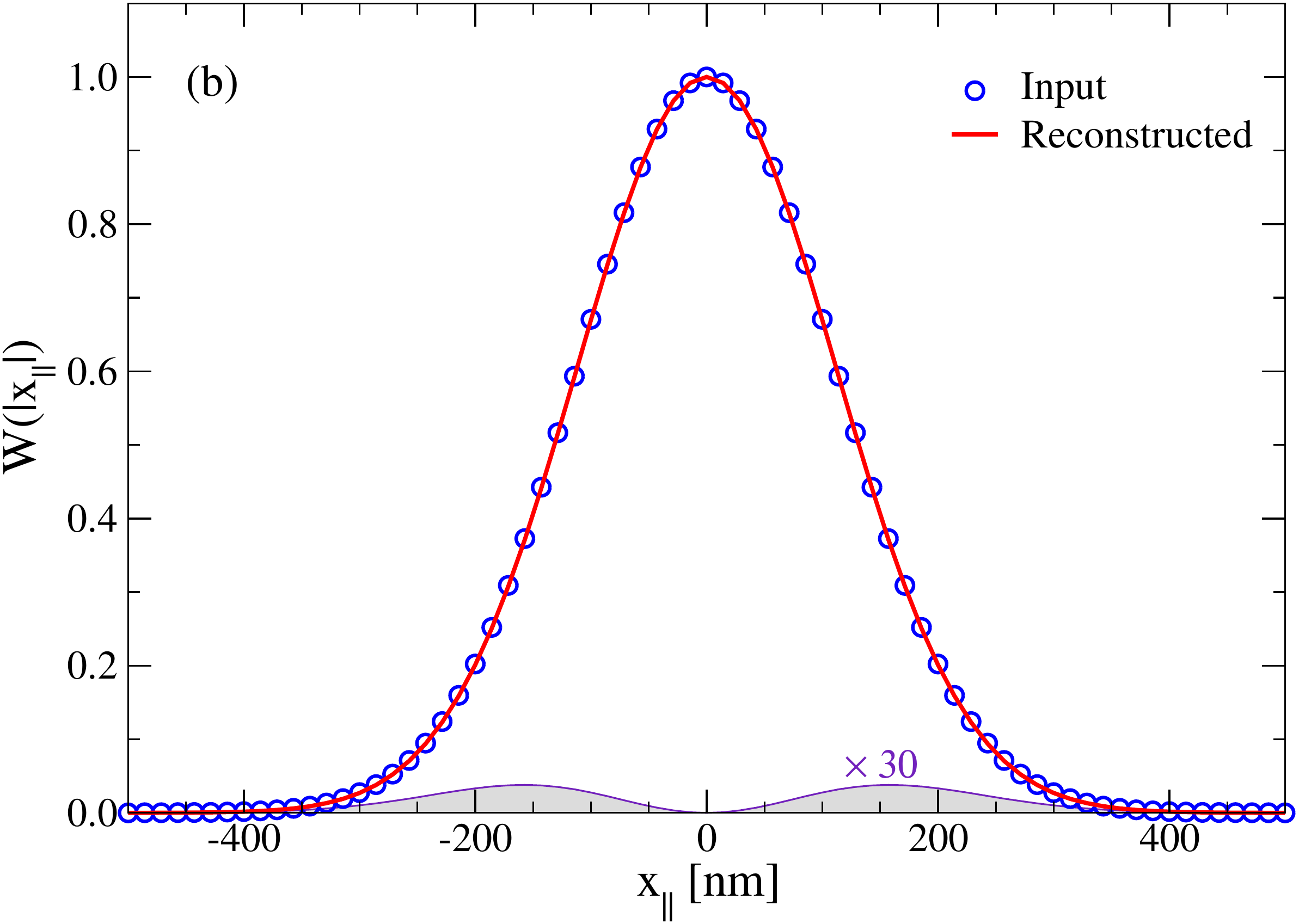} 
  \caption{Same as Fig.~\protect\ref{fig:1}, but now for a Gaussianly correlated surface where the polar angle of incidence is $\theta_0=\ang{50.2}$. The trial function assumed in the reconstruction was the Gaussian form~\eqref{eq:29_Gaussian}, and the values for the parameters obtained were  $\delta^\star=\SI{15.873}{\nano\meter}$ and  $a^\star=\SI{158.000}{\nano\meter}$.
The scattering system assumed in generating the input data were characterized by $\delta=\SI{15.82}{\nano\meter}$, $a=\SI{158.20}{\nano\meter}$, $\e=2.6896$ and $\lambda = \SI{632.8}{\nano\meter}$.}
  \label{fig:Gaussian-1}
\end{figure}

For these parameters, a computer simulation approach~\cite{7} was used to generate scattering data that were obtained by averaging the results from $\num{24000}$  surface realizations. Results obtained this way are presented as open symbols in Fig.~\ref{fig:Gaussian-1}(a) for the polar angle of incidence $\theta_0=\ang{50.2}$. It is these data we will base our inversion on in this subsection, that is, here this data set represents $\la \p R_{ss}/\p\Omega_s\ra_{\textrm{incoh,input}}$. 

Motivated by the reconstruction done in Sec.~\ref{sec:Result-Exponential} using scattering data obtained for the exponentially correlated surface, we will now perform similar inversions for Gaussianly correlated surfaces using various variational parameter sets, ${\cal P}$, that are subsets of $\{ \delta^\star, a^\star, \e^\star, \gamma^\star\}$. In such cases, the starting values assumed in the minimization will be  $\{ \SI{2}{\nano\meter}, \SI{75}{\nano\meter}, 1, 2 \}$, respectively, if nothing is said to indicate otherwise.

%
%
%
Figure~\ref{fig:Gaussian-1} presents the results of the reconstruction for ${\cal P}=\{\delta^\star, a^\star\}$ under the assumption that the trial function used for $W(|\pvec{x}|)$ is of the Gaussian form, Eq.~\eqref{eq:29_Gaussian}. In this way, the reconstruction procedure resulted in the numerical values $\delta^\star=\SI{15.873}{\nano\meter}$ and  $a^\star=\SI{158.000}{\nano\meter}$. These values  agree  rather well with the values assumed in generating the scattering data used in the inversion.  Moreover, the input and reconstructed correlation functions, as well as the absolute difference between them, are depicted in Fig.~\ref{fig:Gaussian-1}(b); the solid red line in Fig.~\ref{fig:Gaussian-1}(a) represents the mean DRC predicted by the inversion. 

%
\begin{table}
  \caption{\label{tab:Gaussian} Summary of the values reconstructed from in-plane scattering data obtained for the polar angle of incidence  $\theta_0=\ang{50.2}$ and corresponding to a Gaussian correlated surface characterized by $\delta=\SI{15.82}{nm}$ and $a=\SI{158.20}{nm}$. The dielectric constant of the substrate was $\varepsilon=\num{2.6896}$. Note that a  Gaussian correlation function corresponds to an exponent $\gamma=\num{2.00}$ for the stretched exponential. 
The symbol ``---'' indicates that the corresponding parameter was not reconstructed and instead had the value assumed in numerically generating the input data. When inverting for any of the parameters in the set $\{ \delta^\star, a^\star, \e^\star, \gamma^\star\}$ the initial values used were $\{ \SI{2}{\nano\meter}, \SI{75}{\nano\meter}, 1, 2 \}$, respectively. 
}
  \begin{ruledtabular}
    \begin{tabular}{lllll}
      $\delta^\star \; \si{[nm]}$ & $a^\star \; \si{[nm]}$ & $\varepsilon^\star$ & $\gamma^\star$  & Comments \quad \\
      \hline 
      \num{15.922} & \num{157.928} & ---         & ---         &  Fig.~\ref{fig:Gaussian-1}    \\
      \num{16.161} & \num{157.785} & \num{2.645} & ---         &  Fig.~\ref{fig:W_Gaussian}(a) \\
      \num{16.170} & \num{154.592} & ---         & \num{1.929} &  Fig.~\ref{fig:W_Gaussian}(b) \\
      \num{16.180} & \num{157.148} & \num{2.651} & \num{1.986} &  Fig.~\ref{fig:W_Gaussian}(c) \\
    \end{tabular}
  \end{ruledtabular}
\end{table}

\begin{figure}[tbph] 
  \centering
  \includegraphics*[width=0.45 \columnwidth]{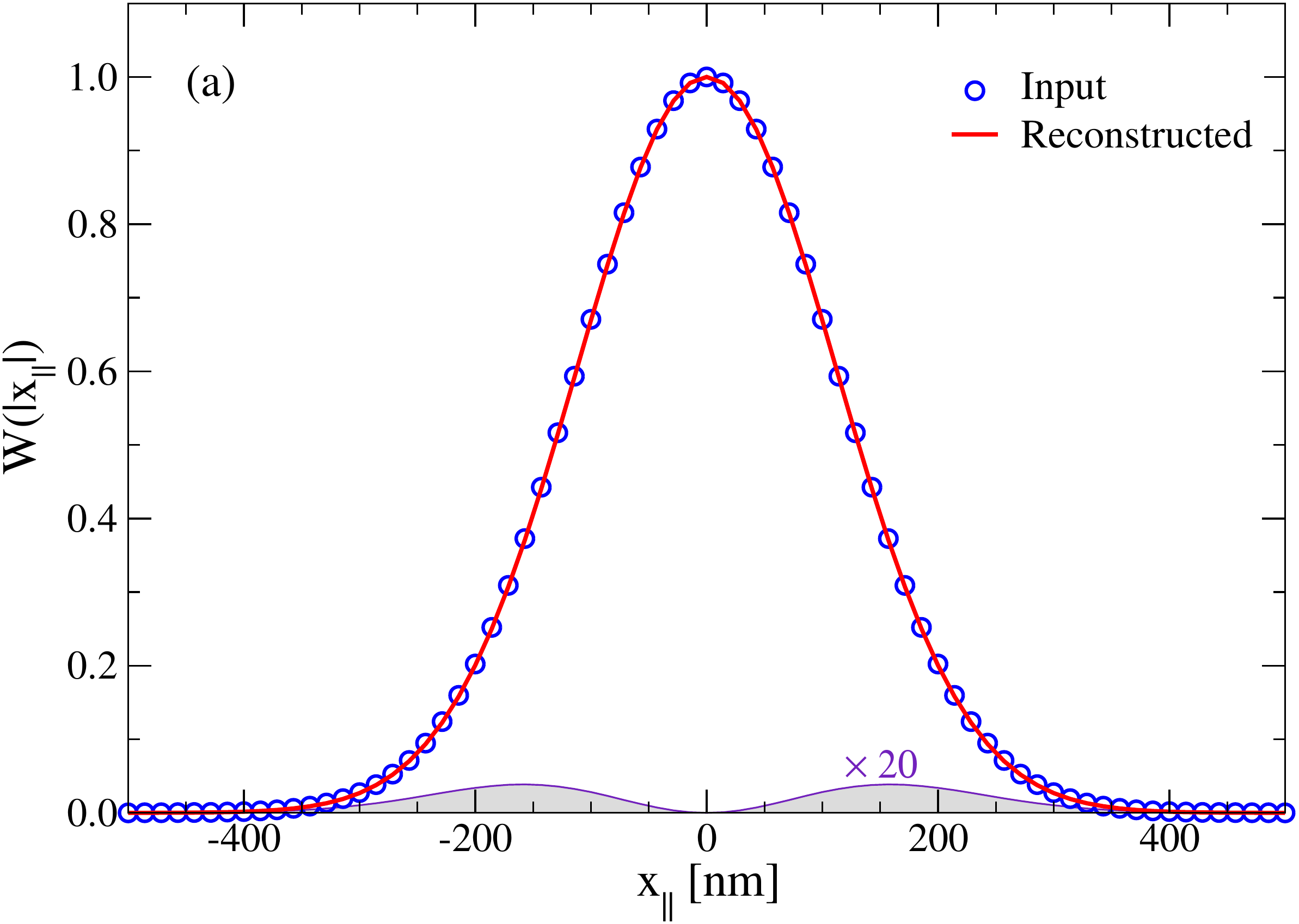} \\
  \includegraphics*[width=0.45 \columnwidth]{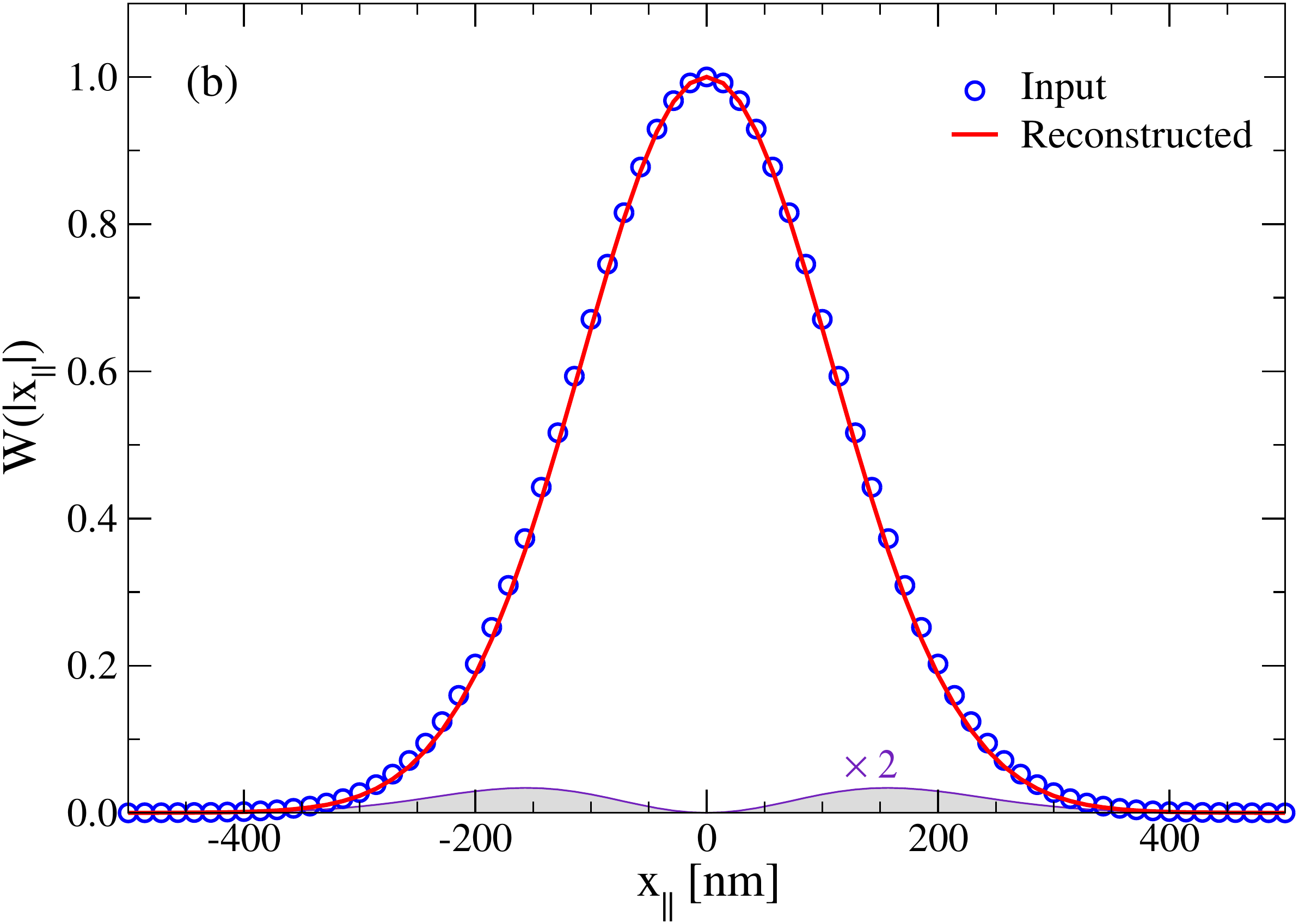} \\
  \includegraphics*[width=0.45 \columnwidth]{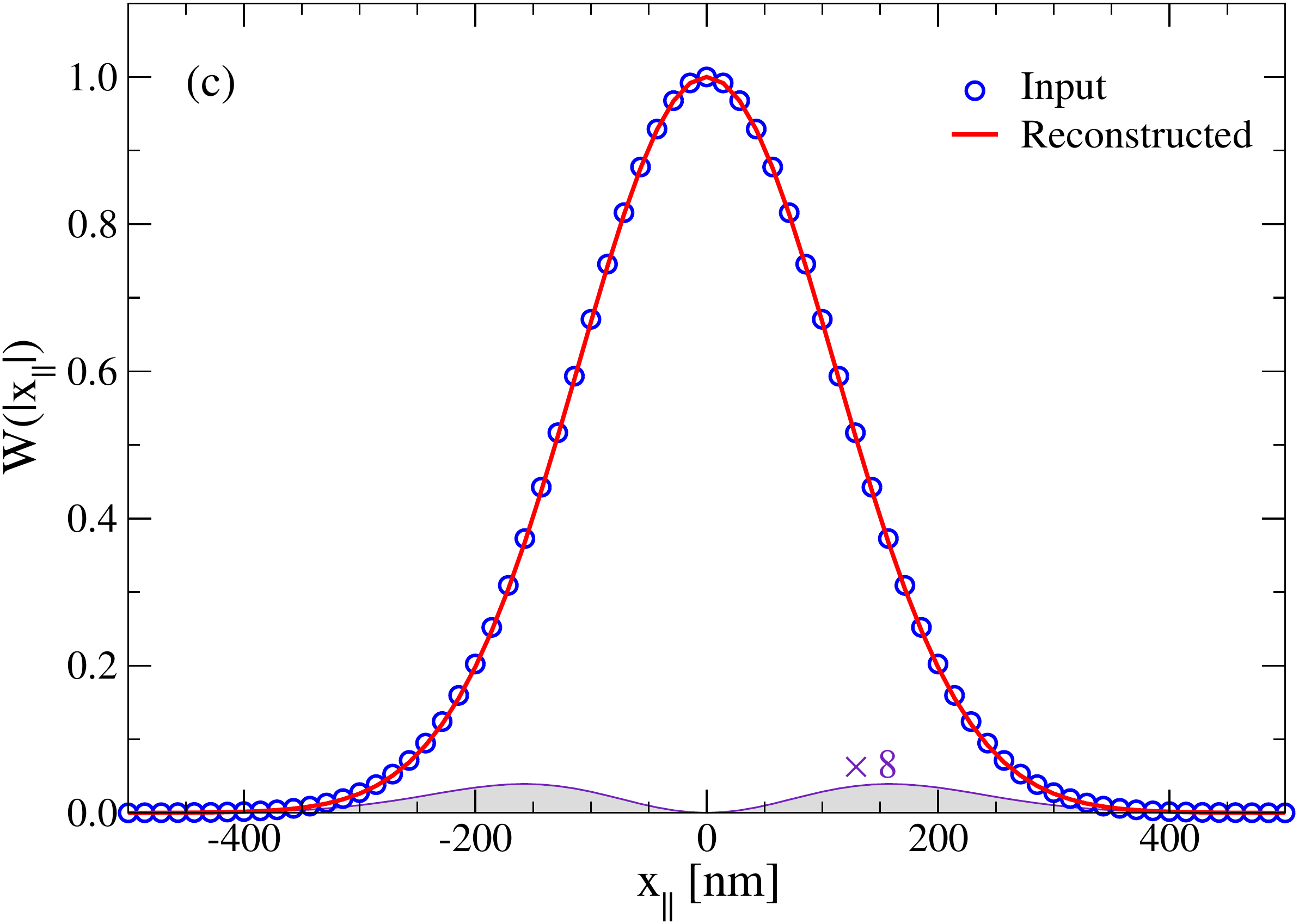} 
\caption{Correlation functions obtained by reconstructing several variational parameter sets ${\cal P}$ for the Gaussian surface roughness parameters defined in Fig.~\protect\ref{fig:Gaussian-1}: 
  (a) ${\cal P}= \{ \delta^\star, a^\star, \e^\star \}$; 
  (b) ${\cal P}= \{ \delta^\star, a^\star, \gamma^\star \}$; 
  and 
  (c) ${\cal P}= \{ \delta^\star, a^\star, \e^\star, \gamma^\star \}$.
  The numerical values of the reconstructed parameters are listed in Table~\ref{tab:Gaussian}. The shaded areas represent the absolute error between the input and reconstructed correlation functions.}
\label{fig:W_Gaussian}
\end{figure}

We now continue by reconstructing the same  variational parameter sets, ${\cal P}$, as were used in Sec.~\ref{sec:Result-Exponential} for the exponential surface roughness; the only difference now is that we will assume the form of the trial function~\eqref{eq:29_Gaussian} where we previously used Eq.~\eqref{eq:29_exponential}. A summary of the reconstructed parameters is presented in Table~\ref{tab:Gaussian}. Moreover, in Fig.~\ref{fig:W_Gaussian} the input and different reconstructed correlation functions obtained in this way are presented, together with comparisons of them. The mean DRC that result from these reconstructions are visually indistinguishable from those of Fig.~\ref{fig:Gaussian-1}(a), and such plots have therefore not been presented.

From Figs.~\ref{fig:Gaussian-1}--\ref{fig:W_Gaussian} and Table~\ref{tab:Gaussian} we see that the reconstructed results for the Gaussianly correlated surface roughness are in general good, at least for the angle of incidence $\theta_0=\ang{50.2}$ assumed here. The quality of the results obtained are on a par with the results obtained previously for the exponentially correlated surface roughness, even though in the Gaussian case the rms-roughness is significantly larger and the angle of incidence is non-zero.

%
%

It should be mentioned that we have also performed reconstructions based on input data corresponding to other polar angles of incidence than $\theta_0=\ang{50.2}$, and it was found that the values of the reconstructed parameters essentially remained unchanged; if there were any changes at all, the quality of the reconstruction seemed to improve for smaller polar angles of incidence. Assuming different initial values of the parameters of the set ${\cal P}$ seemed not to affect the reconstruction. Hence, our results seem to indicate that reconstruction based on different input data, at least for the scattering data that we considered, influences the numerical values of the reconstructed parameters of the surface height correlation function only to a small extent. Furthermore, the reconstruction seems to be reliable for a wide range of angles of incidence.

\subsection{Sensitivity to noise}
\label{sec:Result-Noise}
%

\begin{figure}[tbph] 
  \centering
  \includegraphics*[height=0.33 \columnwidth]{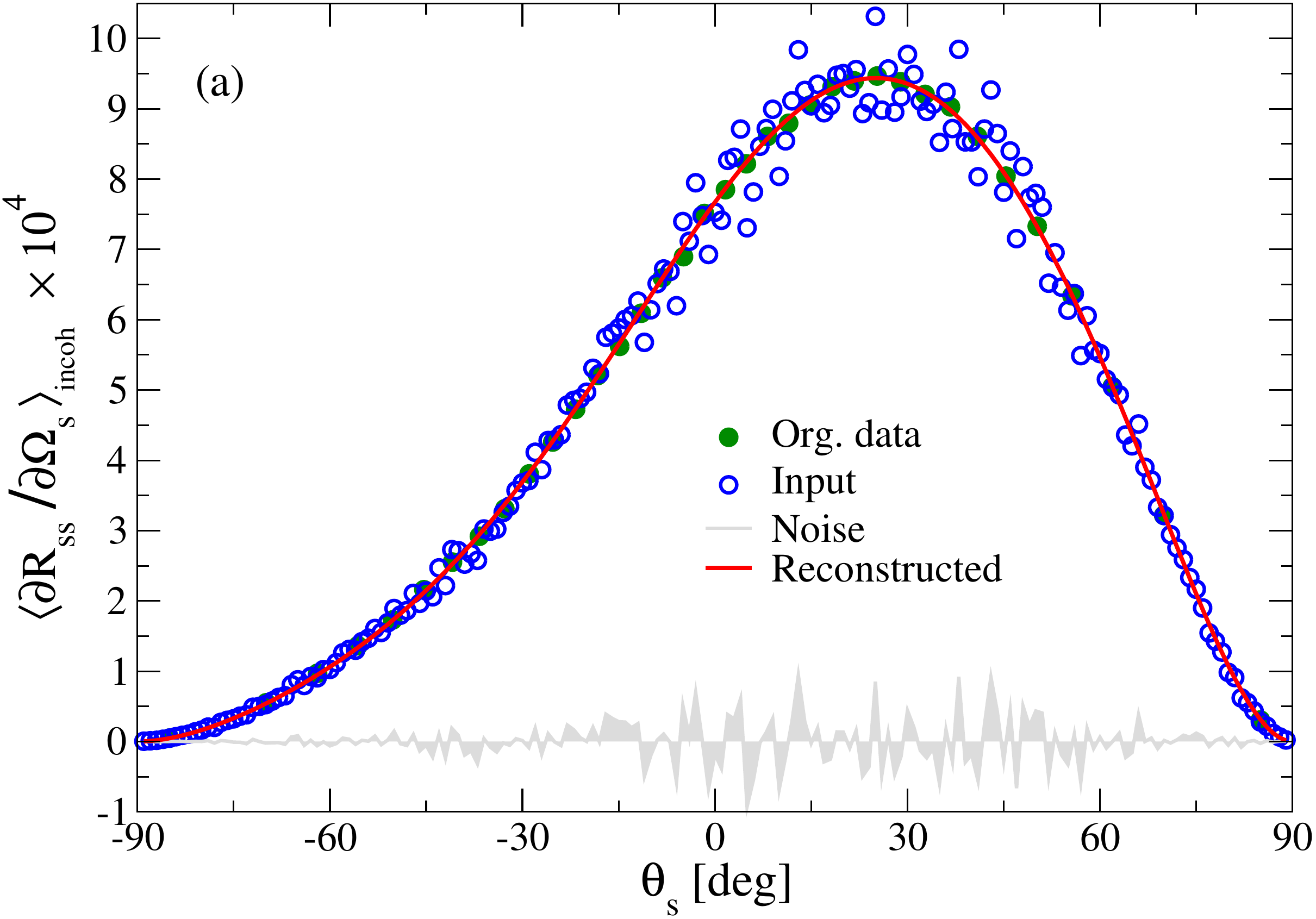}\qquad
  \includegraphics*[height=0.33 \columnwidth]{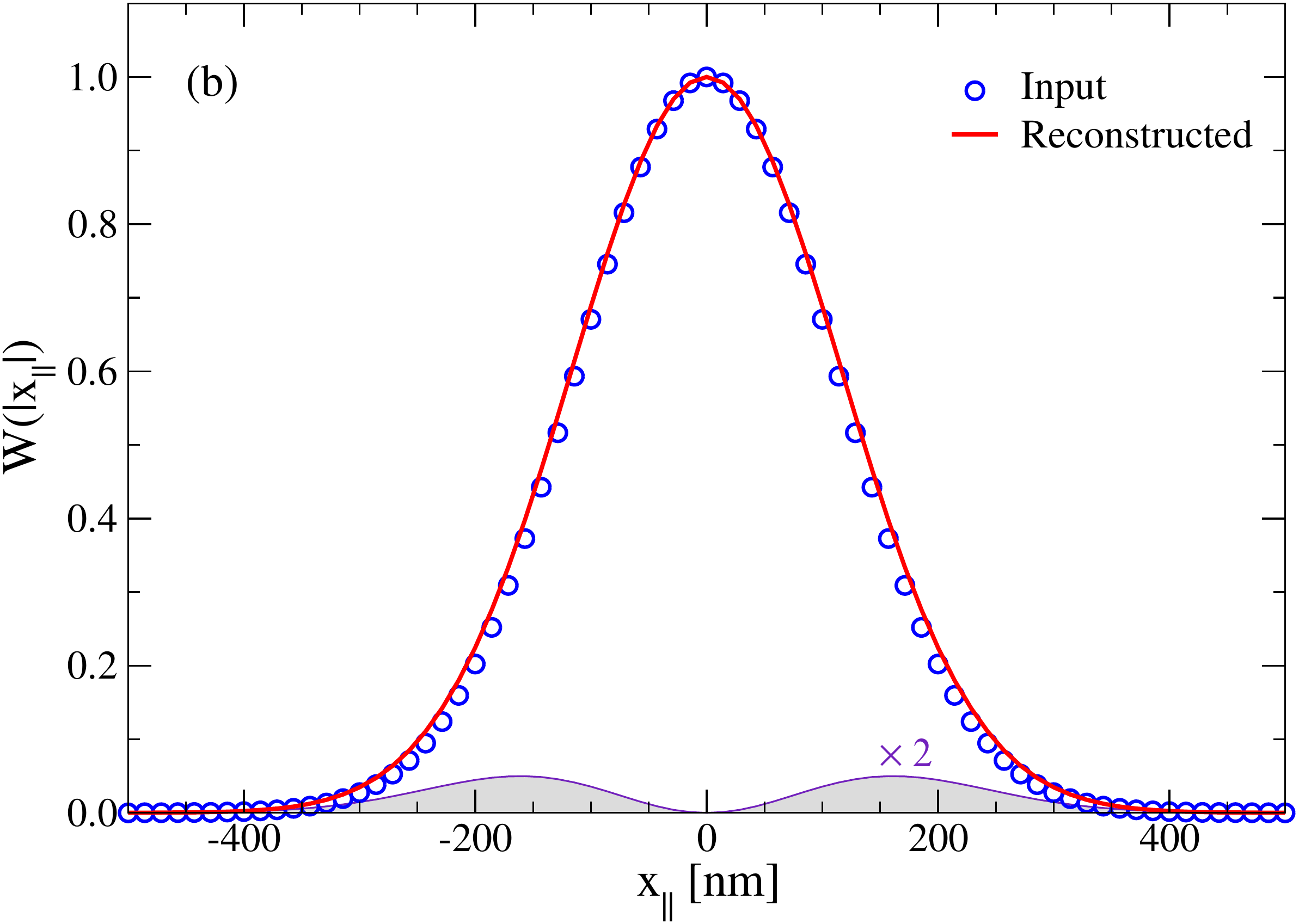}
  \caption{The sensitivity of the reconstruction to multiplicative Gaussian white noise of $5\%$ standard deviation. (a) The incoherent component of the in-plane mean DRC for s-to-s scattering from a Gaussianly correlated surface roughness. The solid symbols represent the same data set that appears as symbols in Fig.~\protect\ref{fig:Gaussian-1}(a). This (smooth) data set was first interpolated to an angular resolution of $\ang{1}$, and then multiplicative Gaussian white noise of $5\%$ standard deviation was added to it. The open blue symbols represent the resulting noisy signal that the reconstruction is based on; the  irregular signal in gray oscillating  around zero is the actual noise being added. The (non-interpolated) original data set is indicated by green filled symbols. The solid red line represents the incoherent component of the mean DRC resulting from reconstructing the variational set ${\cal P}=\{\delta^\star, a^\star, \e^\star,\gamma^\star\}$ (actual values given by the last line of Table~\protect\ref{tab:Gaussian-Noise-5percent}). (b) The input and reconstructed correlation function $W(|\pvec{x}|)$ for variational parameter set ${\cal P}$; the ``non-noisy'' equivalent of this graph is presented in Fig.~\protect\ref{fig:W_Gaussian}(c). It should be noted that reconstruction using any subset of ${\cal P}$ results in more accurate results for the correlation function [see Table~\protect\ref{tab:Gaussian-Noise-5percent}], so what is shown here is indeed the ``worst case'' situation for the cases that we have considered. }
  \label{fig:Noise-5percent}
\end{figure}

Any experimental data set will contain some level of noise. Therefore, it is imperative to have reliable inversion approaches that can be applied successfully to data sets containing noise. Until now,  we have used simulated data as the basis for the reconstruction, and the only source of uncertainty (or noise) in such simulations results is the finite number of surface realizations used to obtain them. However, since a sufficiently high number of realizations has been used in generating such data, the uncertainty has been modest. To start investigating the sensitivity of the reconstructed parameters to noise, we will, for reasons of comparison, base it on the data set used in Sec.~\ref{sec:Result-Gaussian} [open symbols in Fig.~\ref{fig:Gaussian-1}(a)] and add noise to it. Due to the way these simulation results were generated~\cite{7}, only $\num{36}$ points exist in this data set. However, experimental angular resolved scattering data sets typically will have significantly better angular resolution (and, thus, more points). Therefore, to better mimic the more relevant experimental situation, we have interpolated by splines these simulation results [solid symbols in Fig.~\ref{fig:Noise-5percent}(a)] to an angular resolution of $\Delta\theta_s=\ang{1}$ for $\theta_s$  in the interval from $-\ang{90}$ to $\ang{90}$. Then, to these (locally smooth) interpolated data, we have added \textit{multiplicative Gaussian white noise} of a standard deviation of $5\%$ and zero mean (gray erratic signal oscillating around zero in Fig.~\ref{fig:Noise-5percent}(a) resulting in the open blue data in Fig.~\ref{fig:Noise-5percent}(a). It is this latter data set that will be used as the \textit{noisy input signal} for the reconstructions to be performed below.

Results for the reconstructed parameters based on this Gaussian white noisy data set, performed in a fashion identical to what was done in the preceding subsection, are listed in Table~\ref{tab:Gaussian-Noise-5percent}. From this table, we observe that the results obtained are rather good, even for the significant noise level assumed. Moreover, one finds that the quality of the reconstruction is not dramatically degraded compared to what was previously obtained using the  non-noisy data set [see Table~\ref{tab:Gaussian}]. Not surprisingly, the poorest results of the inversion are obtained for the variational parameter set ${\cal P}$ of cardinality \num{4}, and this ``worst case'' is presented as solid red lines in Fig.~\ref{fig:Noise-5percent}. The results presented in Table~\ref{tab:Gaussian-Noise-5percent} and Fig.~\ref{fig:Noise-5percent}, which for the non-noisy case should be compared to Table~\ref{tab:Gaussian} and Fig.~\ref{fig:W_Gaussian}(c), support the view that the reconstruction approach presented in this paper is able to produce reliable results also when the input data are noisy. For instance, when reconstructing \num{4} parameters, the relative error in the reconstructed correlation lengths are about $3.5\%$ and $-0.7\%$ for the noisy and non-noisy case, respectively, which is not dramatic given the level of noise that was added to the input data. 

It should be mentioned that we found that if the parameters were estimated using a noisy version of the original scattering data instead of the interpolated data, as done above, the results using the trial function~\eqref{eq:29_Gaussian} essentially remained unchanged. However, for the stretched exponential form~\eqref{eq:30}, the results were more sensitive to the initial values used in the minimization, resulting in less robust results than those obtained by the use of the former data set. We will see in the next subsection that this can also be the case when using multiple angles of incidence.

%
%

%
\begin{table}
  \caption{\label{tab:Gaussian-Noise-5percent}
    Same as Table~\ref{tab:Gaussian}, but now the data set used in the inversion consisted of an interpolated version of the data set used to produce the results of Table~\ref{tab:Gaussian}, with added multiplicative Gaussian white noise of a standard deviation of $5\%$ (and mean zero). In Fig.~\protect\ref{fig:Noise-5percent}, the noisy input data set is depicted as blue open symbols, while the noise appears in gray.}
  \begin{ruledtabular}
    \begin{tabular}{lllll}
      $\delta^\star \; \si{[nm]}$ & $a^\star \; \si{[nm]}$ & $\varepsilon^\star$ & $\gamma^\star$  & Comments \quad \\
      \hline 
      \num{15.955} & \num{157.705} & ---         & ---         &  --- \\
      \num{16.151} & \num{157.571} & \num{2.653} & ---         &  --- \\
      \num{15.946} & \num{157.824} & ---         & \num{2.003} &  --- \\
      \num{15.941} & \num{163.661} & \num{2.606} & \num{2.102} &  Fig.~\ref{fig:Noise-5percent} \\
    \end{tabular}
  \end{ruledtabular}
\end{table}
\subsection{Inversion of data obtained from multiple angles of incidence}
\label{sec:Result-Multiangle}
%
When the experimental setup is prepared to measure the in-plane scattering of light for one angle of incidence, it is relatively straightforward to perform additional measurements for other angles of incidence. Therefore, it is of interest to study how the reconstructed parameters will depend on using multiple angles of incidence,  and thus several data sets, in the inversion. In order to include, say, $N$ data sets into the reconstruction, the cost function used in the minimization is generalized to 
\begin{align}
  \chi^2({\cal P}) &=  \sum_{n=1}^N \chi^2_n({\cal P}),
  \label{eq:const-func-multiangle}
\end{align}
where $\chi^2_n({\cal P})$ is defined by Eq.~\eqref{eq:27} and corresponds to the contribution to the total cost function $\chi^2({\cal P})$ from a single angle of incidence $\theta_0^{(n)}\in\{\theta_0^{(1)},\theta_0^{(2)},\ldots,\theta_0^{(N)}\}$. 

\begin{figure}[tbph] 
  \centering
  \includegraphics*[height=0.33 \columnwidth]{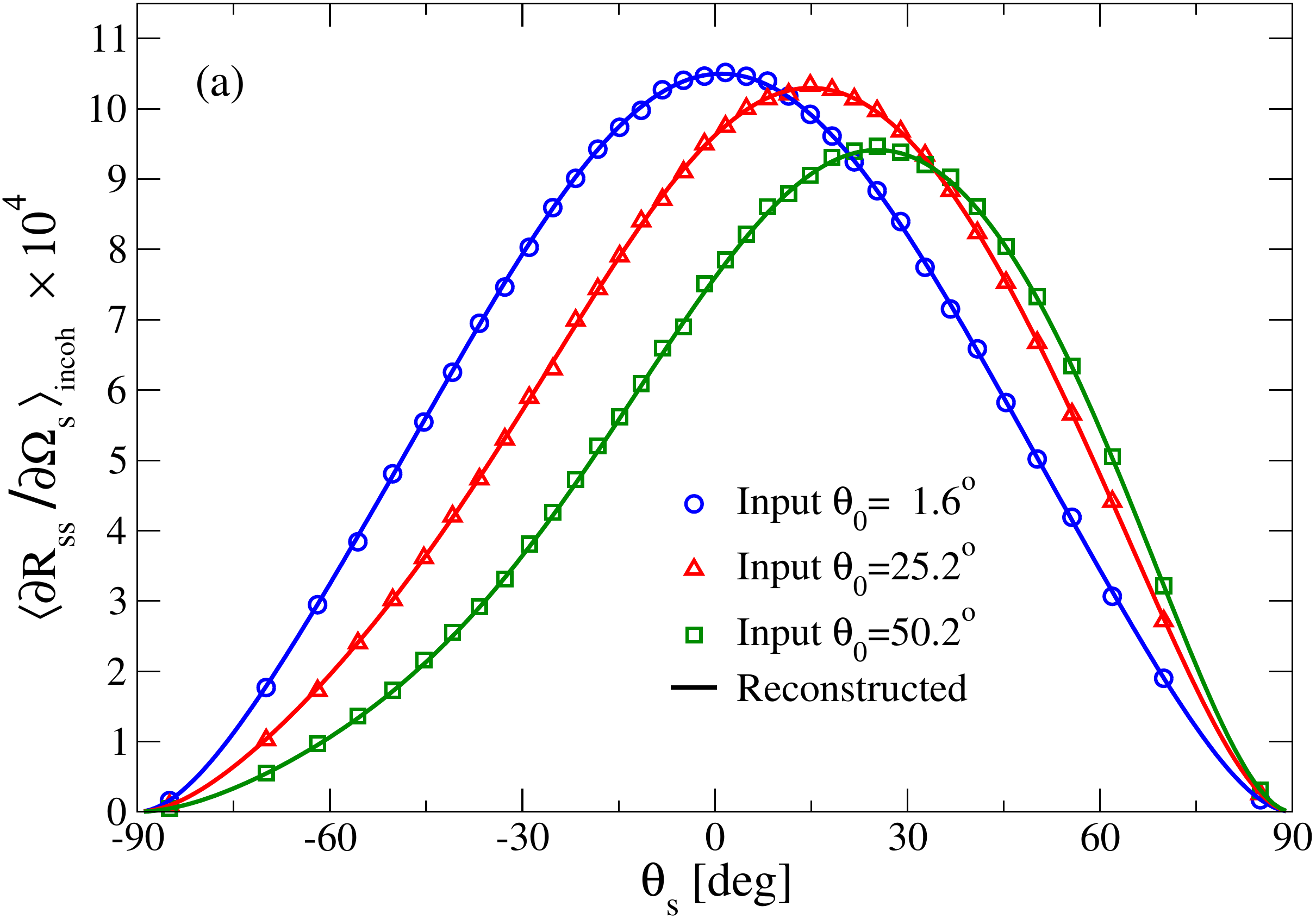}\qquad
  \includegraphics*[height=0.33 \columnwidth]{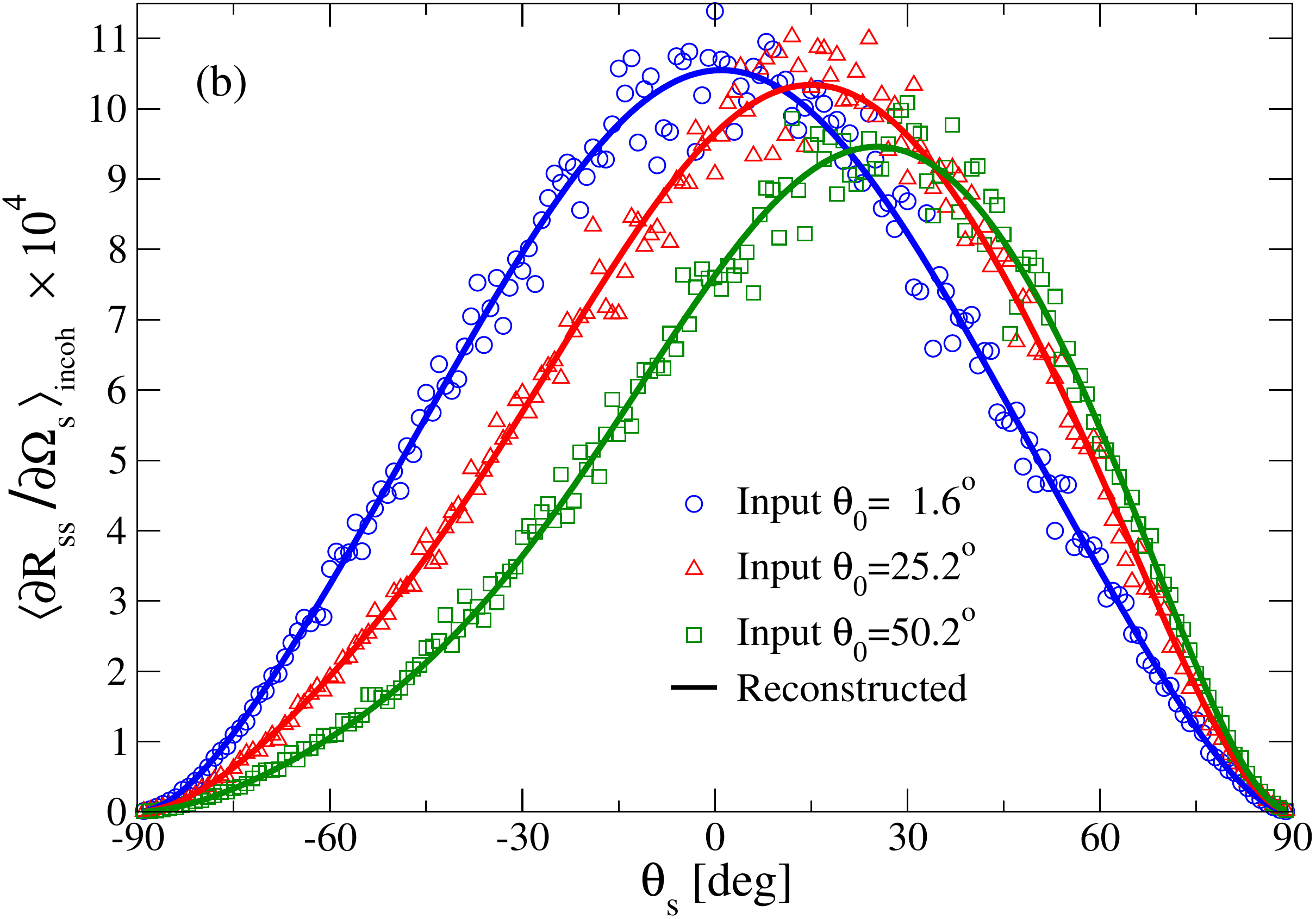}
  \caption{The same as (a) Fig.~\protect\ref{fig:Gaussian-1}(a) and (b) Fig.~\protect\ref{fig:Noise-5percent}(a) but for the polar angles of incidence $\theta_0=\ang{1,6}$, \ang{25.3}, and \ang{50.2} and multi-angle reconstruction of the data sets corresponding to these angles of incidence assuming the variational parameter set ${\cal P}=\{\delta^\star, a^\star, \e^\star\}$. The solid lines, independent of color, represent the reconstructed mean DRCs. The parameter values obtained from the reconstructions are listed in Tables~\protect\ref{tab:Gaussian-multiangles}  and \protect\ref{tab:Gaussian-Noise-5percent-multiangles}. 
}
  \label{fig:multiangle}
\end{figure}

Assuming the Gaussianly correlated surface roughness of the previous subsections, in-plane data for the mean DRCs were obtained from computer simulations for the polar angles of incidence $\theta_0=\ang{1.6}$, \ang{25.3} and \ang{50.2}~\cite{7}. A series of joint inversions were then performed based on the three resulting data sets seen as open symbols in Fig.~\ref{fig:multiangle}(a). The parameters obtained by such an approach are presented in Table~\ref{tab:Gaussian-multiangles} and the resulting mean DRCs obtained when reconstructing $\delta^\star$, $a^\star$ and $\e^\star$ are presented as solid lines in Fig.~\ref{fig:multiangle}(a). A comparison of the results presented in Tables~\ref{tab:Gaussian} and \ref{tab:Gaussian-multiangles} reveals that including additional data sets into the inversion scheme, at least for the data sets we used in obtaining these results, did not change the estimates of the parameters in any significant way. If there is any change, it may seem as if a multi-angle reconstruction may slightly improve the results when using the trial function~\eqref{eq:29_Gaussian}, while it becomes slightly worse for the trial function~\eqref{eq:30}.
However, before one can draw firm conclusions on this issue, proper estimates of the error bars associated with each of these parameters will have to be obtained, something that is outside the scope of the present work.

If now noise is added to these data sets after first interpolating them to a higher angular resolution, as was done above for the single data set corresponding to $\theta_0=\ang{50.2}$, then on performing multi-angle reconstructions based on the resulting data, the values presented in Table~\ref{tab:Gaussian-Noise-5percent-multiangles} are obtained [see also Fig.~\ref{fig:multiangle}(b)]. The first thing to observe from Table~\ref{tab:Gaussian-Noise-5percent-multiangles} is that the multi-angle inversions based on the stretched exponential trial function~\eqref{eq:30} produce rather inaccurate results compared to inversions based on only one of these data sets [see Table~\ref{tab:Gaussian-Noise-5percent}]. Potentially one could first estimate the exponent of the stretched exponential from a single data set, since we have found that it does not matter which of the data sets we use, and then use this value for $\gamma^\star$ as a fixed parameter in a multi-angle inversion. However, we will not consider this situation farther here, since we feel that it is probably more fruitful to consider alternative forms of the trial functions.

On the other hand, when using the Gaussian trial function~\eqref{eq:29_Gaussian} in the minimization, a comparison of the results presented in  Tables~\ref{tab:Gaussian-Noise-5percent} and \ref{tab:Gaussian-Noise-5percent-multiangles} shows  that the multi-angle reconstructions are producing more accurate results than those obtained when the inversion is based on the single angle scattering data set corresponding to $\theta_0=\ang{50.2}$. We have also found this result to be true when any of the other two data sets were used in the single-angle inversion. This is an important result, since it may indicate that including several experimental data sets into the inversion may improve the estimates of the parameters. The results from  Table~\ref{tab:Gaussian-Noise-5percent-multiangles} also hint at the importance of the choice taken for the trial function, since \textit{a priori} it is not known which form of $W(|\pvec{x}|)$ will result in a cost function, $\chi({\cal P})$, that has a deep and well defined minimum, in contrast to many local minima of comparable depths.    

It is also interesting to note that the results obtained for the multi-angle noisy case [Table~\ref{tab:Gaussian-Noise-5percent-multiangles}] seem to be more accurate than those for the corresponding multi-angle non-noisy case [Table~\ref{tab:Gaussian-multiangles}]. However,  here it is important to recall that significantly more points are used in the inversion in the former than in the latter case, and we speculate that this could be the main reason for the improvement.

%
%

%
\begin{table}
  \caption{\label{tab:Gaussian-multiangles} 
    Same as Table~\protect\ref{tab:Gaussian}, but now the reconstruction is based on several data sets corresponding to the  polar angles of incidence $\theta_0=\ang{1.6}$, $\ang{25.3}$ and $\ang{50.2}$ [open symbols in Fig.~\protect\ref{fig:multiangle}(a)].}
  \begin{ruledtabular}
    \begin{tabular}{lllll}
      $\delta^\star \; \si{[nm]}$ & $a^\star \; \si{[nm]}$ & $\varepsilon^\star$ & $\gamma^\star$  & Comments \quad \\
      \hline 
      \num{15.920} & \num{157.649} & ---         & ---         &  --- \\
      \num{16.074} & \num{157.929} & \num{2.658} & ---         &  Fig.~\ref{fig:multiangle}(a) \\
      \num{16.308} & \num{152.620} & ---         & \num{1.896} &  --- \\
      \num{16.208} & \num{155.652} & \num{2.665} & \num{1.953} &  --- \\  
    \end{tabular}
  \end{ruledtabular}
\end{table}

%
\begin{table}
  \caption{\label{tab:Gaussian-Noise-5percent-multiangles} 
    Same as Table~\protect\ref{tab:Gaussian-Noise-5percent}, but now the reconstruction is based on several data sets obtained for the polar angles of incidence $\theta_0=\ang{1.6}$, $\ang{25.3}$ and $\ang{50.2}$. The data sets with the multiplicative $5\%$ standard deviation Gaussian white noise data added to them, on which the reconstructions are based,  are depicted as open symbols in Fig.~\protect\ref{fig:multiangle}(b).}
  \begin{ruledtabular}
    \begin{tabular}{lllll}
      $\delta^\star \; \si{[nm]}$ & $a^\star \; \si{[nm]}$ & $\varepsilon^\star$ & $\gamma^\star$  & Comments \quad \\
      \hline 
      \num{15.843} & \num{158.694} & ---         & ---         &  --- \\
      \num{15.822} & \num{158.863} & \num{2.669} & ---         &  Fig.~\ref{fig:multiangle}(b)  \\
      \num{16.866} & \num{145.591} & ---         & \num{1.749} &  --- \\
      \num{17.050} & \num{140.664} & \num{2.730} & \num{1.675} &  --- \\  
    \end{tabular}
  \end{ruledtabular}
\end{table}

\subsection{Computational cost of the inversion scheme}
%
In principle, an inversion scheme, similar to the one proposed in this work, could be based on one of the rigorous simulation approaches that recently have become available to calculate light scattering from two-dimensional randomly rough surfaces~\cite{our1,our2,review}. However, for such an inversion scheme to be practically relevant, it is has to be numerically efficient, since during the inversion process, the forward scattering problem has to be solved for a large set of parameters. From this perspective, the rigorous numerical simulation approaches mentioned previously fall short since they typically require days of computer time, or more, to run just for one set of parameters. 

In contrast, the phase perturbation theoretical approach to the forward scattering problem that we base our inversion scheme on, is computationally efficient. For instance, for an angular resolution of \ang{1}, it takes less than \SI{2}{\second} to obtain the mean DRC in the plane of incidence for one set of surface parameters (and one angle of incidence). Furthermore, the cpu times it took to perform the four inversions whose results are given in Table~\ref{tab:Gaussian} were \SI{25}{\second} and  \SI{140}{\second} when reconstructing \num{2} and \num{4} parameters, respectively [initial values as in Sec.~\ref{sec:Result-Gaussian}]. The remaining two inversions required cpu times  somewhere in between the two previously given cpu times. All reported computer cpu times are based on the use of a single computer core on an Intel i7 960 CPU operating at \SI{3.20}{\giga\hertz}.
%
%
%
%

Even if such cpu times  do depend strongly on the angular resolution of the data set used in the inversion, the amount of noise that it contains, and the initial values from which the reconstruction is started, these numbers for the computational cost do illustrate that the proposed inversion method is rather quick to perform; this, together with its robustness,  should make it useful in practical situations.

\section{Discussions and Conclusions}
\label{Sec:Conclusions}

In this paper we have shown that second-order phase perturbation theory can be used to model the incoherent scattering of light  by a two-dimensional randomly rough dielectric surface and to calculate the mean differential reflection coefficient for such a surface.  As a result it has been expected that it should be an effective tool for the inversion of light scattering data to obtain statistical properties of a random surface on which the mean differential reflection coefficient depends.  This expectation has been borne out for weakly rough two-dimensional randomly rough dielectric surfaces. Together with several parametrized forms for the normalized surface height autocorrelation function $W(|\bxp |)$, namely an exponential, a Gaussian, and a stretched exponential, and the minimization of a cost function with respect to the parameters defining these forms, phase perturbation theory has been used in this paper to determine $W(|\bxp |)$, the rms height of the surface, the transverse correlation length  of the surface roughness, the dielectric constant of the scattering medium when it was not known in advance, as well as the exponent of the stretched exponential. The function $W (|\bxp |)$ has been reconstructed quite accurately.  The agreement between the reconstructed  values of $\delta$, $a$, and $\e$ and the input values of these parameters is gratifyingly satisfactory.

This agreement remains very good when significant multiplicative Gaussian white noise is included in the input data. When simulated data for multiple  angles of incidence are used in the inversion scheme, it is found that in the absence of noise in the input data the quality of the reconstructions is slightly poorer than when only a single angle of incidence is used. However, when noise is introduced into the input data the use of results obtained from multiple angles of incidence yields slightly better reconstructions than when data from only a single angle of incidence are used. The reasons for this behavior of the reconstructions is not understood, and deserve further study. 

An investigation of the computational cost of the inversion approach developed here shows that it is quite computationally efficient compared to the several orders of magnitude higher computational cost of carrying out the inversions by the use of scattering data obtained by rigorous simulations. The cpu times required for carrying out an inversion calculation using our approach for a given angular resolution, one set of surface parameters, and a single angle of incidence, namely seconds, are short enough that this approach can be useful in practical situations.      

The inversion approach developed here needs to be explored to determine ranges of roughness, wavelength, and dielectric parameters for which it gives reliable results.  Error estimates for the reconstructed values of the roughness and material parameters sought should be obtained.  The use of more flexible trial functions for $W(|\bxp |)$ in reconstructions should be explored, as well as the use of more than one wavelength.  Reflectivity data can serve as the basis of an inversion scheme based on phase perturbation theory, and its utility for this purpose should be examined.  This issue will be explored in subsequent work.

\begin{acknowledgments}
This research was supported in part by NTNU and the Norwegian metacenter for High Performance Computing (NOTUR) by the allocation of computer time.
The research of \O.S.H. and  I.S. was supported in part by The Research Council of Norway Contract No. 216699. 
\end{acknowledgments}


\appendix*
\section{Derivation of Eq. (\ref{eq:14})}


Even if we in this work are concerned with a scattering geometry where the substrate is a dielectric, this assumption will not be made in this Appendix. Instead, the dielectric function of the substrate will here be assumed complex, so that the substrate can be either a dielectric or a metal. This generalization is done in order to facilitate future reference to the results of this Appendix, and because the expressions can be derived simultaneously for a dielectric or metallic substrate with little extra effort.

The starting point for our derivation of Eq.~\eqref{eq:14} is Eqs.~(12), (15), and (16)--(18) of Ref.~\citenum{9}, and the definition of the scattering matrix ${\bf S}(\bqp|\bkp)$ in terms of the matrix of the scattering amplitudes ${\bf R}(\bqp|\bkp)$, Eq.~\eqref{eq:11}. 
From these results we obtain for the $ss$ element of the scattering matrix the expansion
\begin{align}
  \label{eq:A_1}
  S_{ss}(\bqp|\bkp)  &=  S_{ss}^{(0)}(\bqp|\bkp) - \imu S_{ss}^{(1)}(\bqp|\bkp) - \frac{1}{2}S_{ss}^{(2)}(\bqp|\bkp) + \cdots,
\end{align}
where the superscript denotes the order of the corresponding term in the surface profile function $\zxp$. The coefficient $S_{ss}^{(0)}(\bqp|\bkp)$ is given by 
\begin{subequations}
\label{eq:A_2}
 \begin{align}
  \label{eq:A_2a}
  S_{ss}^{(0)}(\bqp|\bkp) 
  & = (2\pi)^2 \delta(\bqp - \bkp)\frac{\alpha_0(\kp) - \alpha(\kp)}{\alpha_0(\kp) + \alpha(\kp)}  \nn \\
  & = (2\pi)^2 \delta(\bqp - \bkp)\left(\frac{\alpha_0(\qp) - \alpha(\qp)}{\alpha_0(\qp) + \alpha(\qp)}\right)^{1/2} 
      \left(\frac{\alpha_0(\kp) - \alpha(\kp)}{\alpha_0(\kp) + \alpha(\kp)}\right)^{1/2}  \nn \\
  & = (2\pi)^2 \delta(\bqp - \bkp)\frac{\left[ \alpha_0^2(\qp) - \alpha^2(\qp)\right]^{1/2}\left[ \alpha_0^2(\kp) - \alpha^2(\kp)\right]^{1/2}}
          {\left[ \alpha_0(\qp) + \alpha(\qp) \right]\left[ \alpha_0(\kp) + \alpha(\kp) \right]}  \nn \\
  & = (2\pi)^2 \delta(\bqp - \bkp)\frac{(1 - \varepsilon)(\w /c)^2}{\left[ \alpha_0(\qp) + \alpha(\qp) \right]\left[ \alpha_0(\kp) + \alpha(\kp) \right]}, 
 \end{align}
where the functions $\alpha_0(\qp)$ and $\alpha(\qp)$ are defined by Eqs.~\eqref{eq:def-alpha0} and \eqref{eq:def-alpha}, respectively.
The coefficient $S_{ss}^{(1)}(\bqp|\bkp)$ is found to be 
 \begin{align}
  \label{eq:A_2b}
  S_{ss}^{(1)}(\bqp|\bkp) 
  & = 2(1 - \varepsilon) \left(\frac{\omega}{c}\right)^2 
  \frac{\alpha_0^{1/2}(\qp)\alpha_0^{1/2}(\kp)}{\left[ \alpha_0(\qp) + \alpha(\qp) \right]\left[ \alpha_0(\kp) + \alpha(\kp) \right]}
  (\hbqp \cdot \hbkp) \hat{\zeta}(\bqp - \bkp),
\end{align}
while the coefficient $S_{ss}^{(2)}(\bqp|\bkp)$ is given by 
\begin{align}
  \label{eq:A_2c}
  S_{ss}^{(2)}(\bqp|\bkp) 
    =& - \frac{4 \alpha_0^{1/2}(\qp)\alpha_0^{1/2}(\kp)}{\left[ \alpha_0(\qp) + \alpha(\qp) \right]\left[ \alpha_0(\kp) + \alpha(\kp) \right]}  \nn \\
  &  \qquad \times \Big\{-\frac{1}{2}(1 - \varepsilon) \left(\frac{\omega}{c}\right)^2 [\alpha(\qp) + \alpha(\kp)](\hbqp \cdot \hbkp) \hat{\zeta}^{(2)}(\bqp - \bkp)    \nn \\
  &  \qquad\qquad + \frac{(1 - \varepsilon)^2}{\varepsilon} \left( \frac{\w}{c}\right)^2 \int \dfint[2]{\pp}{(2 \pi)^2} \hat{\zeta}(\bqp - \bpp) (\hbqp \times \hbpp)_3 \alpha(\pp) (\hbpp \times \hbkp)_3 \hat{\zeta}(\bpp - \bkp) \Big\}\nn \\ 
  & + \frac{4 \alpha_0^{1/2}(\qp)\alpha_0^{1/2}(\kp) (1 - \varepsilon)(\w/c)^2}{\left[ \alpha_0(\qp) + \alpha(\qp) \right]\left[ \alpha_0(\kp) + \alpha(\kp) \right]} 
 \int \dfint[2]{\pp}{(2 \pi)^2} \hat{\zeta}(\bqp - \bpp)  \nn \\
  &  \qquad  \times
\Bigg\{ \frac{1 - \varepsilon}{\varepsilon}(\hbqp \times \hbpp)_3 
\frac{\alpha^2(\pp)}{\varepsilon \alpha_0(\pp) + \alpha(\pp)}(\hbpp \times \hbkp)_3  
+ (1 - \varepsilon) \left( \frac{\w}{c} \right)^2 
\frac{(\hbqp \cdot \hbpp)(\hbpp \cdot \hbkp)}{\alpha_0(\pp) + \alpha(\pp)}
    \Bigg\} \hat{\zeta}(\bpp - \bkp)  \nn \\
   =& - 4 (1 - \varepsilon) \left(\frac{\omega}{c}\right)^2 
 \frac{\alpha_0^{1/2}(\qp)\alpha_0^{1/2}(\kp)}{\left[ \alpha_0(\qp) + \alpha(\qp) \right]\left[ \alpha_0(\kp) + \alpha(\kp) \right]} 
 \nn \\   
& \qquad \times \int \dfint[2]{\pp}{(2 \pi)^2} \hat{\zeta}(\bqp - \bpp) 
    \Bigg\{ - \frac{1}{2}[\alpha(\qp) + \alpha(\kp)](\hbqp \cdot \hbkp) + 
\frac{1 - \varepsilon}{\varepsilon} (\hbqp \times \hbpp)_3 \alpha(\pp) (\hbpp \times \hbkp)_3  \nn \\
& \qquad \quad \quad  
 - \frac{1 - \varepsilon}{\varepsilon} (\hbqp \times \hbpp)_3 \frac{\alpha^2(\pp)}{\varepsilon \alpha_0(\pp) + \alpha(\pp)} (\hbpp \times \hbkp)_3 
-(1 - \varepsilon)\left(\frac{\omega}{c}\right)^2  \frac{(\hbqp \cdot \hbpp)(\hbpp \cdot \hbkp)}{\alpha_0(\pp) + \alpha(\pp)} 
    \Bigg\} 
     \hat{\zeta}(\bpp - \bkp)
\nn \\
   =& - 4 (1 - \varepsilon) \left(\frac{\omega}{c}\right)^2 
 \frac{\alpha_0^{1/2}(\qp)\alpha_0^{1/2}(\kp)}{\left[ \alpha_0(\qp) + \alpha(\qp) \right]\left[ \alpha_0(\kp) + \alpha(\kp) \right]} 
 \nn \\   
& \quad\; \times \int \dfint[2]{\pp}{(2 \pi)^2} \hat{\zeta}(\bqp - \bpp) 
    \Bigg\{ - \frac{1}{2}[\alpha(\qp) + \alpha(\kp)](\hbqp \cdot \hbkp) 
    + 
    (1 - \varepsilon)(\hbqp \times \hbpp)_3 
    \frac{ \alpha_0(\pp)\alpha(\pp) }{ \varepsilon \alpha_0(\pp) + \alpha(\pp)}
    (\hbpp \times \hbkp)_3  \nn \\
& \qquad \qquad \qquad  \qquad \qquad  \qquad
 -(1 - \varepsilon)\left(\frac{\omega}{c}\right)^2  \frac{(\hbqp \cdot \hbpp)(\hbpp \cdot \hbkp)}{\alpha_0(\pp) + \alpha(\pp)} 
    \Bigg\} 
     \hat{\zeta}(\bpp - \bkp).
\end{align}
\end{subequations}
In obtaining this expression we have used the result that 
\begin{align}
  \label{eq:A_3}
  \hat{\zeta}^{(2)}(\bqp - \bkp)  = \int \dfint[2]{\pp}{(2 \pi)^2} \hat{\zeta}(\bqp - \bpp) \hat{\zeta}(\bpp - \bkp). 
\end{align}

When we substitute the results given by Eqs.~\eqref{eq:A_2} into Eq.~\eqref{eq:A_1} we find that through terms of second order in the surface profile function, $S_{ss}(\bqp|\bkp)$ takes the form 
 \begin{align} \label{eq:A_4}
   S_{ss}(\bqp|\bkp)  
   =&
   \sgn(\pvecUnit{q}\cdot \pvecUnit{k}) 
   \frac{(1 - \varepsilon)(\w/c)^2}{\left[ \alpha_0(\qp) + \alpha(\qp) \right]\left[ \alpha_0(\kp) + \alpha(\kp) \right]}  \nn \\
   &  \times 
      \Bigg\{
       (2\pi)^2 \delta(\bqp - \bkp) \sgn(\pvecUnit{q}\cdot \pvecUnit{k}) 
       - 2\imu\alpha_0^{1/2}(\qp)\alpha_0^{1/2}(\kp) 
       | \pvecUnit{q}\cdot \pvecUnit{k} | 
        \hat{\zeta}(\bqp - \bkp)  \nn \\
   &  \qquad -2 \alpha_0^{1/2}(\qp)\alpha_0^{1/2}(\kp) \sgn(\pvecUnit{q}\cdot \pvecUnit{k}) 
        \int \dfint[2]{\pp}{(2 \pi)^2} \hat{\zeta}(\bqp - \bpp) 
        \bigg[ \frac{1}{2}\left[\alpha(\qp) + \alpha(\kp) \right](\hbqp \cdot \hbkp)  \nn \\
   & \qquad \quad + (\varepsilon - 1) (\hbqp \times \hbpp)_3 \frac{\alpha_0(\pp) \alpha(\pp)}{\varepsilon \alpha_0(\pp) + \alpha(\pp)} (\hbpp \times \hbkp)_3  
           -(\varepsilon - 1) \left( \frac{\w}{c} \right)^2 \frac{(\hbqp \cdot \hbpp)(\hbpp \cdot \hbkp)}{\alpha_0(\pp) + \alpha(\pp)} 
              \bigg]  \hat{\zeta}(\bpp - \bkp) \Bigg\}.
\end{align}
This expression for $S_{ss}(\bqp|\bkp)$ is manifestly reciprocal, i.e.\ it satisfies Eq.~\eqref{eq:12b}.Moreover, for reasons of later convenience, in writing Eq.~\eqref{eq:A_4} we have factored out a phase $\sgn(\pvecUnit{q}\cdot \pvecUnit{k})$, where $\sgn(\cdot)$ denotes the sign function defined by $x=\sgn(x)|x|$. 

We next express Eq.~\eqref{eq:A_4} in the form of a Fourier integral:
\begin{align}
  \label{eq:A_5}
  S_{ss}(\bqp|\bkp)
  & =
  \sgn(\pvecUnit{q}\cdot \pvecUnit{k})
  \frac{(1 - \varepsilon)(\w/c)^2}{d_s(\qp) d_s(\kp)} \int \dint[2]{\xp} \exp{\left[-\imu(\bqp - \bkp)\cdot \bxp \right]}   
  \bigg\{1 - 2\imu\alpha_0^{1/2}(\qp)\alpha_0^{1/2}(\kp)
  |\hbqp \cdot \hbkp |   \zeta(\bxp) \nn\\   
  & \qquad -2\alpha_0^{1/2}(\qp)\alpha_0^{1/2}(\kp)
   \int \dfint[2]{\pp}{(2 \pi)^2}  
  F(\bqp|\bpp|\bkp)  
  \int \dint[2]{\up}  
  \exp{\left[-\imu(\bpp - \bkp)\cdot \bup\right]} \zeta(\bxp) \zeta(\bxp + \bup) \bigg\}, 
\end{align}
where 
\begin{align}
 \label{eq:A_6}
 F(\bqp|\bpp|\bkp) 
  =&
   \sgn(\pvecUnit{q}\cdot \pvecUnit{k}) 
   \bigg\{  
  \frac{1}{2} \left[\alpha(\qp) + \alpha(\kp) \right](\hbqp \cdot \hbkp) + 
 (\varepsilon - 1) (\hbqp \times \hbpp)_3 \frac{\alpha_0(\pp) \alpha(\pp)}{d_p(\pp)} (\hbpp \times \hbkp)_3  
 \nn \\
 & \qquad  \qquad \qquad 
 -(\varepsilon - 1) \left(\frac{\w}{c} \right)^2 
          \frac{(\hbqp \cdot \hbpp)(\hbpp \cdot \hbkp)}{d_s(\pp)} 
   \bigg\},
\end{align}
and the functions $d_p(p_\parallel)$ and $d_s(p_\parallel)$ are defined in Eq.~\eqref{eq:15}. One notes from Eq.~\eqref{eq:A_6} that $F(\bqp|\bpp|\bkp) = F(-\bqp|\bpp|\bkp)$ so that the expression inside the curly brackets in Eq.~\eqref{eq:A_5} is a continuous function of the lateral scattering wave vector $\pvec{q}$ (and in particular at $\pvec{q}=\vec{0}$).

From Eq.~\eqref{eq:A_5} we find that 
\begin{subequations}
  \label{eq:A_7}
\begin{align}
 \langle S_{ss}(\bqp|\bkp) \rangle 
 & =     
   \sgn(\pvecUnit{q}\cdot \pvecUnit{k})
    \frac{(1 - \varepsilon)(\w/c)^2}{d_s(\qp) d_s(\kp)}
     \int \dint[2]{\xp} \exp{\left[-\imu(\bqp - \bkp)\cdot \bxp \right]}  \nn \\
 & \qquad \times 
 \bigg\{1 - 2 \delta^2\alpha_0^{1/2}(\qp)\alpha_0^{1/2}(\kp) 
     \int \dfint[2]{\pp}{(2 \pi)^2} F(\bqp|\bpp|\bkp) g(|\bpp - \bkp|) \bigg\}   
     \label{eq:A_7_a}   
  \\ 
& \cong    
   \sgn(\pvecUnit{q}\cdot \pvecUnit{k}) 
       \frac{(1 - \varepsilon)(\w/c)^2}{d_s(\qp) d_s(\kp)}
           \int \dint[2]{\xp} \exp{\left[-\imu(\bqp - \bkp)\cdot \bxp\right]}  \nn \\
& \qquad \times \exp\left[ 
                           - 2 \delta^2\alpha_0^{1/2}(\qp)\alpha_0^{1/2}(\kp)
                            \int \dfint[2]{\pp}{(2 \pi)^2} 
                            F(\bqp|\bpp|\bkp) g(|\bpp - \bkp|)
                    \right].
    \label{eq:A_7_b}
\end{align}
\end{subequations}

It follows that 
\begin{align}
 \label{eq:A_8}
\left|\left< S_{ss}(\bqp|\bkp) \right> \right|^2 
  & =  
     \left| \frac{(1 - \varepsilon)(\w/c)^2}{ d_s(\qp) d_s(\kp)}    \right|^2  
     \exp{\left[-2M(\bqp|\bkp)\right]}
          \int \dint[2]{\xp} \int \dint[2]{x_\parallel'} \exp{\left[-\imu(\bqp - \bkp)\cdot (\bxp - \bxpp)\right]}, 
\end{align}
where
\begin{align}
 \label{eq:A_9}
 2M(\bqp|\bkp) 
 & = 4 \delta^2\alpha_0^{1/2}(\qp)\alpha_0^{1/2}(\kp) 
     \,\Re \!\!  \int \dfint[2]{\pp}{(2 \pi)^2} F(\bqp|\bpp|\bkp) g(|\bpp - \bkp|).
\end{align}

We next find that 
\begin{align}
 \label{eq:A_10}
 \left\langle \left| S_{ss}(\bqp|\bkp) \right|^2  \right\rangle 
  & = 
       \left| \frac{(1 - \varepsilon)(\w/c)^2}{ d_s(\qp) d_s(\kp)}    \right|^2   
      \int \dint[2]{\xp} \int \dint[2]{x_\parallel'} \exp{\left[-\imu(\bqp - \bkp)\cdot (\bxp - \bxpp) \right]} \nn \\
 & \quad \times \Big\{1 - 2\imu\alpha_0^{1/2}(\qp) \alpha_0^{1/2}(\kp) 
                    |\hbqp \cdot \hbkp|  \left<\zeta(\bxp)- \zeta(\bxpp)\right> 
            + 4 \alpha_0(\qp) \alpha_0(\kp)(\hbqp \cdot \hbkp)^2 
                \left< \zeta(\bxp)\zeta(\bxpp) \right>  \nn \\
 & \qquad \quad -2 \alpha_0^{1/2}(\qp) \alpha_0^{1/2}(\kp) 
           \int \dfint[2]{\pp}{(2 \pi)^2}   \int \dint[2]{\up} 
           \Big[\exp{\left[-\imu(\bpp - \bkp)\cdot \bup\right]}   F(\bqp|\bpp|\bkp)
            \left< \zeta(\bxp)\zeta(\bxp + \bup) \right>  \nn \\
 & \qquad \qquad \quad  
               + \exp\left[\imu(\bpp - \bkp)\cdot \bup\right] F^*(\bqp|\bpp|\bkp)
                 \left< \zeta(\bxpp)\zeta(\bxpp + \bup) \right>    \Big] \Big\}. 
\end{align}

From this result we obtain
\begin{align} \label{eq:A_11}
 \left\langle \left| S_{ss}(\bqp|\bkp) \right|^2 \right\rangle 
  =& 
     \left| \frac{(1 - \varepsilon)(\w/c)^2}{ d_s(\qp) d_s(\kp)}    \right|^2  
       \int \dint[2]{\xp}  \int \dint[2]{x_\parallel'} \exp{\left[-\imu(\bqp - \bkp)\cdot (\bxp - \bxpp) \right]} \nn \\
 & \quad \times \left[1 + 4\delta^2 \alpha_0(\qp) \alpha_0(\kp)(\hbqp \cdot \hbkp)^2 W(|\bxp - \bxpp|) - 2M(\bqp|\bkp)\right] 
  \nn \\
  \cong& 
        \left| \frac{(1 - \varepsilon)(\w/c)^2}{ d_s(\qp) d_s(\kp)}    \right|^2   
        \exp\left[-2M(\bqp|\bkp)\right] 
        \int \dint[2]{\xp} \int \dint[2]{x_\parallel'} \exp{\left[-\imu(\bqp - \bkp)\cdot (\bxp - \bxpp)\right]} \nn \\  
 & \quad \times 
     \exp\left[ 4 \delta^2\alpha_0(\qp)\alpha_0(\kp)(\hbqp \cdot \hbkp)^2 W(|\bxp - \bxpp|)  \right].
\end{align}

Thus, we have finally
\begin{align}
 \label{eq:A.12}
\Big< \left| S_{ss}(\bqp|\bkp) \right|^2 \Big> - & \left| \Big< S_{ss}(\bqp|\bkp) \Big> \right|^2 
  =
   S \left| \frac{(1 - \varepsilon)(\w/c)^2}{ d_s(\qp) d_s(\kp)}    \right|^2  
  \exp[-2M(\bqp|\bkp)] \nn \\
  & \quad 
 \times \int \dint[2]{\up} \exp{[-\imu(\bqp - \bkp)\cdot \bup] } 
 \left\{\exp\left[4 \delta^2\alpha_0(\qp)\alpha_0(\kp)(\hbqp \cdot \hbkp)^2 W(| \pvec{u} |)\right] - 1 \right\}.
\end{align}
The substitution of this result into Eq.~\eqref{eq:13} yields Eq.~\eqref{eq:14}.


\end{document}